\documentclass[twocolumn]{aastex62}
\usepackage{amsmath}
\usepackage[super]{nth}
\usepackage{fontawesome}

\newcommand{\ts}{\textsuperscript}

\graphicspath{{./}{figures/}}

\published{\today}
\begin{document}

\title{Multiband Probabilistic Cataloging: A Joint Fitting Approach to Point-source Detection and Deblending}

\shorttitle{Multiband Probabilistic Cataloging}
\shortauthors{Feder et al.}

\author{Richard M. Feder}
\affiliation{California Institute of Technology Division of Physics, Math, and Astronomy.
1200 East California Boulevard, Pasadena, CA 91125}

\author{Stephen K. N. Portillo}
\affiliation{DIRAC Institute, Department of Astronomy, University of Washington, 3910 15th Ave NE, Seattle, WA 98195}

\author{Tansu Daylan}
\affiliation{MIT Kavli Institute for Astrophysics and Space Research. 77 Massachusetts Avenue, 37-241, Cambridge, MA 02139}

\author{Douglas Finkbeiner}
\affiliation{Department of Physics, Harvard University, Cambridge, MA}

\begin{abstract}

  Probabilistic cataloging (PCAT) outperforms traditional cataloging methods on single-band optical data in crowded fields.  We extend our work to multiple bands, achieving greater sensitivity ($\sim$ 0.4 mag) and greater speed (500x) compared to previous single-band results.  We demonstrate the effectiveness of multiband PCAT on mock data, in terms of both recovering accurate posteriors in the catalog space and in directly deblending sources.  When applied to Sloan Digital Sky Survey (SDSS) observations of M2, taking \emph{Hubble Space Telescope} data as truth, our joint fit on $r$- and $i$-band data goes $\sim0.4$ mag deeper than single-band probabilistic cataloging and has a false discovery rate less than 20\% for F606W$\leq 20$.  Compared to DAOPHOT, the two-band SDSS catalog fit goes nearly 1.5 mag deeper using the same data, and maintains a lower false discovery rate down to F606W$\sim 20.5$. Given recent improvements in computational speed, multiband PCAT shows promise in application to large-scale surveys and is a plausible framework for joint analysis of multi-instrument observational data. \textbf{  \href{https://github.com/RichardFeder/multiband_pcat}{\faGithub}} 

\end{abstract}


\section{Introduction} \label{sec:intro}
Astronomical catalogs condense the information from observations that can be used to study the properties and evolution of astrophysical sources in detail. The information from catalogs can also be used to perform precise measurements that test fundamental physics. For example, catalogs that map the spatial distribution of galaxies in the universe (e.g. Sloan Digital Sky Survey [SDSS], Dark Energy Survey, etc.), enable the study of the large scale structure that constrains models of dark matter and dark energy \citep{EISENSTEIN}. 

The process of converting intensity maps of observed photons into a catalog of sources with measured properties is inherently lossy. Information such as source-source covariance is often discarded, either because it has a demonstrably negligible impact on the quantity of interest or because assuming so makes subsequent analysis simpler. As modern telescopes have become more sensitive to incoming light, the quality of astronomical data products has improved dramatically, enabling astronomers to make more precise measurements and pursue more sophisticated analyses. At the same time, many assumptions that were once robust in the cataloging process have become less justifiable as a result of these developments. One such assumption is that point sources in an image do not overlap with one another.  While this may be reasonable for the brightest sources in an image, the abundance of newly observed fainter sources makes blending increasingly prevalent. Ground-based telescopes will suffer from blending because they have the increased flux sensitivity to pick up fainter sources, but not the angular resolution to disentangle them from other faint neighbors. One report on the observing strategy for the Large Synoptic Survey Telescope (LSST) notes that spatial confusion will set the absolute scale on several figures of merit \citep{LSST}. A recent report on software strategy for the Hyper-Suprime Cam (HSC) noted that ``58\%
of objects in the HSC Wide survey are blended, in the sense that they were detected as part of an above-threshold region of sky containing multiple significant peaks in surface brightness'' \citep{BOSCH}. Even space telescopes such as the \emph{Wide Field Infrared Space Telescope} (WFIRST) will be plagued by crowding issues, since blended stars and galaxies significantly degrade photometric redshift estimation \citep{WFIRST}. Left unaddressed, blending issues will significantly degrade the science potential of upcoming astronomical surveys.  

In the context of source parameter estimation, maximum likelihood techniques are not well equipped to handle crowded regions or those with low signal-to-noise ratio (S/N). A catalog that maximizes a prescribed likelihood function will still suffer from systematic errors. In the crowded limit, the notion of a maximum likelihood catalog becomes ill-defined, since a point-source model may describe an image as well as or better than another with fewer sources. Because conventional cataloging approaches only produce one realization of the catalog, they are unable to probe the space of alternative catalogs that are consistent with the data through these degeneracies. One might expand the hypothesis space of catalogs by reducing the flux threshold for source detection; however, this can lead to overestimating the number of point sources, since a catalog with a large number of model sources can be fit arbitrarily well to data without providing any meaningful physical description of that data. Even in the limit of infinite data, the maximum likelihood catalog would fit noise peaks. There is often a trade-off between underfitting by missing real sources and overfitting by including spurious ones.

These challenges have inspired the development of probabilistic cataloging, which aims to retain more information from complex datasets and appropriately propagate uncertainties from source-source covariance, sky background, and other nuisance parameters, even in the low-S/N limit. Rather than deriving a single catalog from an image, probabilistic cataloging (PCAT) generates a catalog ensemble, in which each catalog is sampled from the inferred posterior catalog distribution.  In crowded fields, source parameter uncertainties are covariant with those of neighboring sources in a way that manifests in uncertainty on the number of sources. The hypothesis space explored is the union of models containing different numbers of sources. By exploring this space, PCAT generates a fair ensemble that retains comprehensive uncertainties on both source parameters and source number. A catalog ensemble also enables one to propagate deblending uncertainties to downstream analyses. 

While many cataloging methods impose a detection threshold to discriminate between high- and low-confidence sources, PCAT benefits from the ability to propose subthreshold sources. While one may not be interested in the measurement of low-confidence sources, one may want to characterize the effect of these sources on other emission components. If one wants to estimate, for example, background levels coming from specific astrophysical processes in a way that relies on source masking, a catalog ensemble allows one to marginalize this estimate over the contributions from low-confidence sources that might immediately be discarded as background fluctuations. Likewise, one may want the brightest sources to have errors marginalized over the effects of fainter neighboring sources.  

A great deal of work has been done developing probabilistic cataloging into a viable method for use in next-generation surveys like those of LSST and \emph{WFIRST}. Since its first implementation in \cite{BREWER}, probabilistic cataloging has been developed to analyze optical \citep{PORTILLO_17}, X-ray \citep{JONES_15}, and gamma ray observations \citep{DAYLAN_1}. In particular, \cite{PORTILLO_17} demonstrated that, using the same $r$-band SDSS image of M2 and comparing to a \emph{Hubble Space Telescope} (HST) catalog of the same region, PCAT goes more than a magnitude deeper than traditional catalogers while maintaining a lower false discovery rate brighter than \nth{20} magnitude. 

In this paper we present an updated version of PCAT that is significantly faster than that from \cite{PORTILLO_17}, and incorporates the joint analysis of multiband datasets. By simultaneously fitting multiple observations, one can improve point-source inference as a direct result of increased S/N. Furthermore, color information provides additional benefits missed by analysis on stacked observations. A large number of astrophysical sources are best understood through analysis of their spectral emission. There is evidence that spectral information in probabilistic point-source inference can disentangle emissions from highly blended sources. Through simulations of two neighboring X-ray sources with different spectra, \cite{JONES_15} demonstrated that probabilistic catalog inference using spectral information performed better than spatial-only models at favoring a two-source description, in contrast to a single combined source/spectrum. The peaked nature of X-ray source spectra also makes it easier to distinguish sources from the X-ray background, which has a flatter spectrum.

\cite{DAYLAN_1} use a similar implementation of multiband probabilistic cataloging to study gamma ray emission in the north Galactic polar cap. That being said, the corresponding method for optical data operates somewhat differently. One notable difference is that the optical point spread function (PSF; taking SDSS as an example) is on the arcsecond scale, nearly three orders of magnitude smaller than the \emph{Fermi}-LAT PSF \citep{FERMI_LAT}. The higher spatial resolution in optical allows us to treat each image as an equally spaced Cartesian grid over which we use a fixed, Lanczos-interpolated PSF template for cataloging. The use of a PSF template makes the task of probabilistic cataloging computationally simpler in a number of ways, but it also means that the quality of the catalog ensemble is highly dependent on the quality of the template. 

More generally, there are a number of challenges that must be addressed to ensure the success of multiband cataloging in the optical regime:
\begin{enumerate}
\item \emph{Positional Calibration}. In order to make effective proposals in multiple bands, pixel coordinate transformations across bands need to be quick and precise. Astrometric solutions specified using World Coordinate System (WCS) conventions are typically used to do these transformations, but when tasked with evaluating millions of transformations during probabilistic cataloging, a naive application of WCS transformations is not fast enough (see \S 5). The precise mapping to celestial coordinates typically requires the evaluation of trigonometric functions and high-order polynomial distortion coefficients, all of which are computationally expensive. 
\item \emph{Color Priors}. Within the Bayesian framework of probabilistic cataloging, color information offers a well-motivated way to inform the inference of sources and reduce the model parameter space explored by PCAT's catalog ensemble. However, overly restrictive color priors make it difficult to discover astrophysical sources in unique areas of color-color space. Furthermore, the priors used in catalog inference will be folded into any downstream analyses using a probabilistic catalog. As such, it is important that the likelihood can be recovered by investigators who may have different prior beliefs \citep{HOGG_18}.

\item \emph{Computational Speed}. Any application of probabilistic cataloging to survey-scale datasets needs to be fast and scalable.

\end{enumerate}
This work addresses several of these practical challenges. Tests on mock datasets demonstrate the effectiveness of multiband PCAT at both at recovering source populations (\S 3) and at disentangling highly blended sources (\S 4). After presenting a method for cross-observation astrometric calibration that is both precise and much faster than naive application of existing astrometric transformation packages (\S 5), we demonstrate the advantages of multiband PCAT on SDSS photometric data (\S 6).

\section{Methods} \label{sec:methods}
\subsection{Generative Model and Likelihood}
Within the framework of probabilistic cataloging, catalogs are best thought about as samples drawn from the model space -- many of the challenges of cataloging are common to inference in high-dimensional spaces. We define a \textit{catalog} as a sample containing $N$ sources $\{(x_n, y_n, f_n, ...)\}_{n=1}^N$ describing some data. We only consider point sources in this paper, so our catalog will typically look like $\{(x_n, y_n, f_{1,n}, f_{2,n}, ..., f_{B,n})\}_{n=1}^N$, where $\{f_{b,n}\}_{b=1}^B$ represent the fluxes in $B$ different bands for source $n$ and $(x_n,y_n)$ denotes the source position in a fiducial image. We assume the astrometric transformations to $(x,y)$ in other images to be known and fixed. Alternative catalogs that include extended sources such as galaxies require more complicated parameterizations to describe the orientation and morphology of each source. 

Adapting notation from \cite{PORTILLO_17}, we express the expected counts at each pixel grid coordinate $(x,y)=(l,m)$ for band $b$ as $\lambda_{lm}^b$, with units of photoelectrons. This is calculated as a sum of the background in band $b$, $I_{sky}^b$, and the contributions of nearby sources:
\begin{equation}
\lambda_{lm}^b = I_{sky}^b + \sum_{n=1}^N f_{b,n} \mathcal{P}_b(l-x_{b,n}, m-y_{b,n})
\end{equation}
In the above equation, $\mathcal{P}_b(\Delta x, \Delta y)$ is the pixel-convolved PSF extracted by the standard SDSS pipeline for the center of our image. For our observations, $\lambda_{lm}^b$ is large enough that the expected noise is approximately Gaussian with a standard deviation of $\sqrt{\lambda_{lm}^b}$ photoelectrons. For data $k_{lm}^b$ over $n_b$ images with width $W$ and height $H$, the likelihood is 
\begin{equation}
\mathcal{L} = \prod_{b=1}^{B}\prod_{l=1}^W\prod_{m=1}^H \frac{1}{\sqrt{2\pi\lambda_{lm}^b}}\exp\left(-\frac{(k_{lm}^b-\lambda_{lm}^b)^2}{2\lambda_{lm}^b}\right).
\label{likelihood}
\end{equation}
For the purpose of MCMC sampling, we calculate the log-likelihood, which turns our products into sums over pixels and bands:
\begin{align}
\log \mathcal{L} &= \sum_{b=1}^{B}\sum_{l=1}^w\sum_{m=1}^H -\frac{1}{2}\log \lambda_{lm}^b +c - \frac{(k_{lm}^b-\lambda_{lm}^b)^2}{2\lambda_{lm}^b} \\
&\approx \sum_{b=1}^{B}\sum_{l=1}^w\sum_{m=1}^H - \frac{(k_{lm}^b-\lambda_{lm}^b)^2}{2\lambda_{lm}^b}.
\end{align}
When sampling, we only need to compute delta log-likelihoods $\Delta \log \mathcal{L}$, so we are at liberty to discard the additive constant $c$. While we assume a Gaussian likelihood on the pixels of the images, we choose to also remove the logarithmic term $-\frac{1}{2}\log \lambda_{lm}^b$ when computing $\Delta \log \mathcal{L}$. Because we use $\lambda_{lm}^b$ as an estimator of the pixel-level variance as shown in Equation \eqref{likelihood}, an imperfectly estimated PSF will misestimate source fluxes and thus $\lambda_{lm}^b$, meaning that our estimate of the pixel-level variance will be incorrect. Furthermore, the logarithmic term is more costly to compute and has little effect compared to the last term as long as $\lambda_{lm}^b \gg 1$. The range of $\lambda_{lm}^b$ will depend on the background and noise level -- in the case where the background level is high enough and the noise is small compared to the background, the above condition should hold.


The Bayesian element of catalog inference allows us to impose a number of priors on the parameters of interest. These priors, used in combination with the log-likelihood during sampling, are outlined in Appendix \ref{A}.
\subsection{Hierarchical Modeling}
\begin{figure*}
    \centering
    \includegraphics[width=0.4\linewidth]{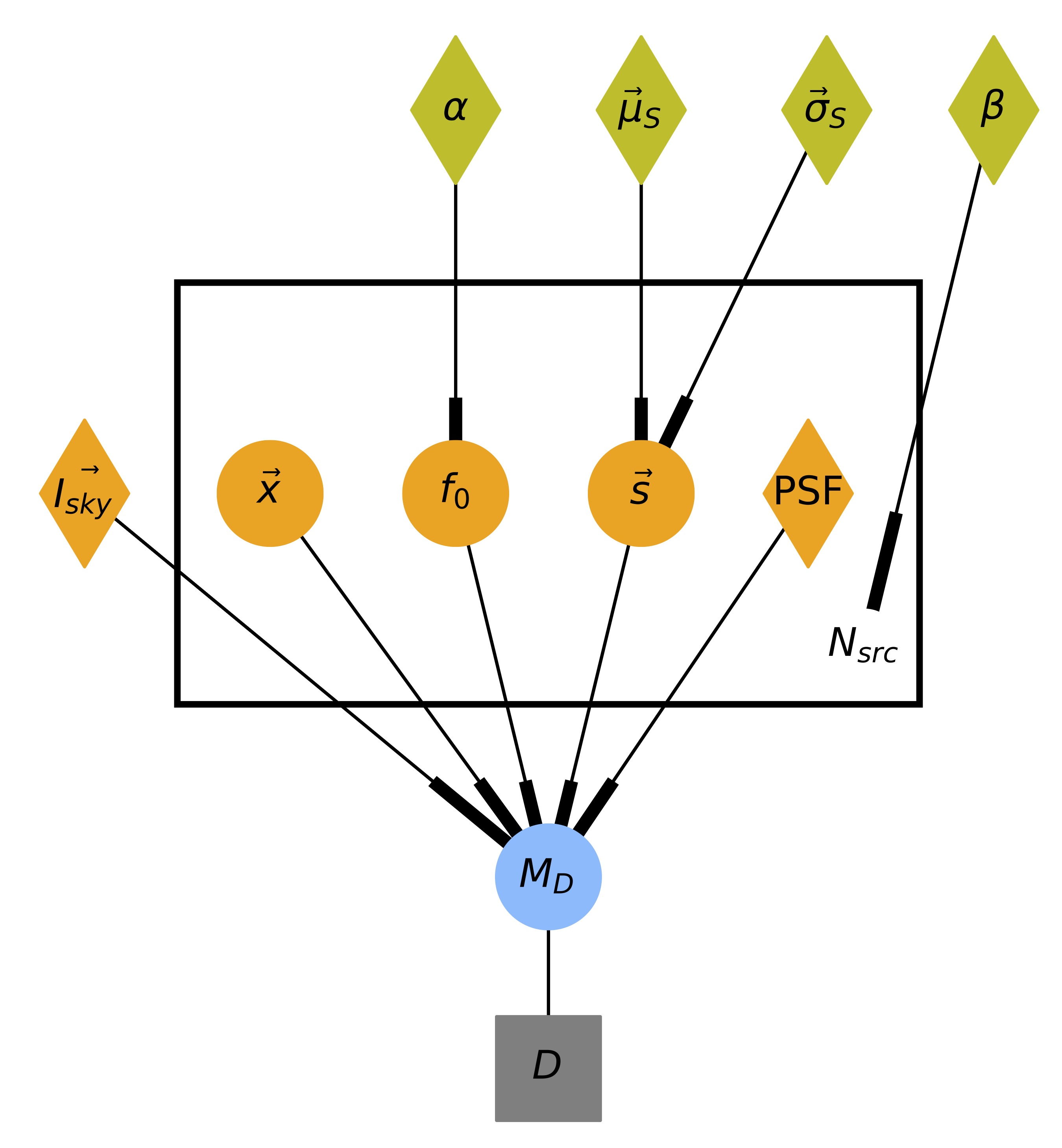}
    \caption{Probabilistic graphical model for multiband PCAT. Each colored node represents a parameter or set of parameters in the model. Vectorized parameters are modeled across multiple bands. Nodes within the plate are replicated $N_{src}$ times and correspond to individual source parameters. $M_D$ refers to the full model, while $D$ refers to the observed photometric data. Circles and diamonds indicate variables that are floated and fixed, respectively.}
    \label{fig:pgm}
\end{figure*}

From the assumption that all point sources in our model are independently realized and neither spatially nor spectrally correlated with each other, one can construct a hierarchical Bayesian model. In a hierarchical model, each point source has priors on its parameters (e.g. flux, color), which are then conditioned by \textit{hyperparameters} that can also float as free parameters in the fit. These hyperparameters might, for example, parameterize the flux distribution, color distribution, or spatial distribution of a source population. Constraints on these hyperparameters (i.e. \emph{hyperpriors}) are analogous in function to priors on source parameters and are part of what makes the model hierarchical. When calculating marginalized estimates on catalog source properties, one can also marginalize over hyperparameter uncertainties. Hierarchical models can be abstracted to any level, with hyperparameters of hyperparameters \emph{ad infinitum}, though in practice the usefulness of models with multiple levels of hyperparameters depends on the quality of data. Nonetheless, the hierarchical Bayesian framework of probabilistic cataloging exploits the ability to propagate prior knowledge from any number of parameters to later parts of the analysis, where those parameters can be constrained with more comprehensive uncertainties. A probabilistic graphical model of our hierarchical model is shown in Figure \ref{fig:pgm}. The hyperparameters in our model are fixed in the following analysis, though in principle they can be floated. 

\subsection{Transdimensional Sampling}
One issue that arises when cataloging an image concerns the fact that the number of unknowns (i.e., the number of sources) is an unknown itself. From a modeling perspective, the challenge lies in the fact that each source has a number of associated parameters, so models with varying numbers of sources have varying dimensionality. Furthermore, sampling across models of different dimension ($\mathcal{D}(x) \neq \mathcal{D}(x')$) would not be bijective, meaning that the Markov chain would not satisfy detailed balance. A solution to this issue is provided by \cite{GREEN} through a modified algorithm known as Reversible Jump MCMC (RJ-MCMC). RJ-MCMC samples across models by drawing random auxiliary variables $u$ and $u'$ with densities $g(u)$ and $g(u')$ to match the dimensions of the initial and final states, where \begin{equation} \mathcal{D}(x) + \mathcal{D}(u) = \mathcal{D}(x') + \mathcal{D}(u') \end{equation} and $\mathcal{D}$ is the dimension operator \citep{DAYLAN_1, GREEN}. In traversing between model spaces of different dimension, $x'$ is determined by $x, u$ and $x$ is determined by $x', u'$. So long as the mapping $(x, u) \leftrightarrow (x', u')$ is a diffeomorphism (i.e., bijective and differentiable in both directions), one can construct proposals so that RJ-MCMC has the same convergence properties as traditional MCMC methods. In the context of probabilistic cataloging, these auxiliary variables are related to model sources being added or removed (often referred to as ``births" and ``deaths"). For example, in the case of a birth, $u$ contains the position and flux of the new source, and $u'$ is the empty set. Sources may also be split or merged together with similar auxiliary variables.       


While one can explore the parameter space of a catalog model with fixed dimension, the transdimensional nature of probabilistic cataloging enables us to explore a generalized \textit{catalog space}, which we define as the union of fixed-dimensional catalog parameter spaces. This is written as
\begin{equation}
\mathcal{C} = \bigcup\limits_{N=N_{min}}^{N_{max}} \mathcal{C}_N = \bigcup\limits_{N=N_{min}}^{N_{max}} X_N \times Y_N \times \mathcal{F}_N \times ...
\label{eqn:catalogunion}
\end{equation} where $N$ is the number of sources in a given catalog model. So long as detailed balance is satisfied, samples drawn from the stationary distribution of the Markov chain inherently capture covariances across models of differing dimensionality (differing numbers of sources) by exploring the posterior distribution of $N$. This enables direct, statistically consistent inference on the number of observed sources in an image.

Our implementation of probabilistic cataloging uses Metropolis-Hastings sampling to converge on the stationary distribution. We provide the acceptance factors computed for various proposals in Appendix \ref{B}.  

\subsection{Computational Requirements}
The code used in this work is an extension of \texttt{Lion}, a significantly faster implementation of PCAT. Here we describe a few of the performance improvements in the updated implementation.

The most computationally expensive part of each step in the Markov chain is constructing and evaluating the model image for a proposed catalog. For a $100\times 100$ pixel single-band image in a crowded field, the implementation of PCAT used in \cite{PORTILLO_17} took $\sim 10^9$ model evaluations over 12 CPU-days to converge. That implementation evaluates the model image by looping over sources and then looping over pixels, using bilinear interpolation on the PSF to determine the contribution of the current source to the current pixel. \texttt{Lion} loops over catalog difference entries instead, saving significant computational time. However, constructing the model image remains the dominant computational cost.

To speed up model evaluation, \texttt{Lion} rewrites PSF subpixel shifts as a matrix multiplication operation, which allows us to take advantage of highly optimized libraries such as CBLAS.\footnote{\url{https://www.gnu.org/software/gsl/doc/html/cblas.html}} Consider the model image for a point source $i$ at location $(x_i, y_i)$. We would like to compute the model image in a postage stamp around the pixel containing the source, choosing a postage stamp large enough to contain the flux from the source but not so large as to introduce needless computation. For each pixel $j$ in the postage stamp, the pixel-convolved PSF must be evaluated at the center of the pixel $(x_j, y_j)$:
\begin{equation}
    \lambda_{i,j} = f_i \mathcal{P}(x_j - x_i, y_j - y_i).
\end{equation}
The PSF can be described by functions for each pixel in the postage stamp that only depend on $(\Delta x_i, \Delta y_i)$, the separation between the source and the center of the pixel it is in:
\begin{equation}
    \lambda_{i,j} = f_i \mathcal{P}_j(\Delta x_i, \Delta y_i).
\end{equation}
These functions $\mathcal{P}_j$ are approximated as third-degree polynomials in $\Delta x$ and $\Delta y$:
\begin{multline}
    \mathcal{P}_j(\Delta x, \Delta y) \approx q_{j,1} + q_{j,2}\Delta x + q_{j,3}\Delta y + q_{j,4}\Delta x^2 \\ + q_{j,5}\Delta x \Delta y + q_{j,6}\Delta y^2 + q_{j,7}\Delta x^3 \\ + q_{j,8}\Delta x^2 \Delta y + q_{j,9}\Delta x \Delta y^2
    + q_{j,10}\Delta y^3.
\end{multline}
For a fixed PSF, the coefficients $q_{j,k}$ can be precomputed. Then, if a matrix $\textbf{Q}$ is constructed with $\textbf{Q}_{jk} = q_{j,k}$ and the column vector
\begin{multline}
    \textbf{X}_i = f_i \times \large(1, \Delta x_i, \Delta y_i, \Delta x_i^2, \Delta x_i\Delta y_i, \Delta y_i^2, \\ \Delta x_i^3, \Delta x_i^2 \Delta y_i, \Delta x_i \Delta y_i^2, \Delta y_i^3\large)
\end{multline}
is constructed for the source, the postage stamp for the source (flattened into a column vector) is simply the matrix product $\textbf{QX}_i$. Collecting the column vectors for all sources $\textbf{X}_i$ into a single matrix $\textbf{X}$, all of the postage stamps are columns in the matrix product $\textbf{QX}$. Once this matrix product is computed, the postage stamps can be extracted and pasted onto the larger model image. This procedure agrees with bilinear interpolation to $\sim 0.1\%$, and takes less than 1 ms for 500 sources in a $100\times 100$ pixel image, compared to 100 ms for the previous implementation.

In addition to speeding up model evaluation, tuning the step sizes and evaluating multiple proposals with one model image makes each step in the chain more efficient so that only $10^6$ steps are required. Given the same single-band image, the updated implementation takes only half an hour on one CPU to converge, over 500 times faster than the previous implementation. Other than its drastically improved run time, this implementation of probabilistic cataloging shares many of the advantages and disadvantages of the previous implementation. More information about the updated implementation, \texttt{Lion}, will be detailed in a future paper. The code used in this work, \texttt{Lion 0.1}, is publicly available on GitHub \href{https://github.com/RichardFeder/multiband_pcat}{\faGithub}.

\section{Mock Data}

In this section we present the results of multiband PCAT applied to mock simulated datasets. By comparing our multiband catalog ensemble to true mock catalog parameters realized from a fully generative model, we can (1) confirm that multiband PCAT converges appropriately and (2) understand the relative improvement in catalog ensemble fidelity as a function of the number of bands used.

\subsection{Mock data generation and results}
We first generate a mock SDSS dataset in three bands ($r$, $i$, $g$). Our point sources have $r$-band fluxes drawn from a flux distribution with a power-law slope $\alpha= -2$, after which we draw colors $(r-i) \sim \mathcal{N}(0.25, 0.5)$ and $(g-r)\sim \mathcal{N}(0.25, 0.5)$ to calculate $i$- and $g$-band fluxes. We fix the background levels in our bands to $\vec{b}=(b_r, b_i, b_g)=(179, 310, 115)$ in analog-digital units (ADU). These values are based off of empirical background estimates from SDSS run 2583, camcol 4, field 136, which we will use in \S 6. We fix the PSF in all bands to the SDSS $r$-band PSF template from the same field. This PSF is used to generate our mock catalog sources, and is assumed to be known perfectly in our mock tests. For our first mock runs, we assume the astrometric solution of each source to be perfect across all bands by setting $(x,y)_r = (x, y)_i = (x, y)_g$. Once the model image is generated in each band, we add Gaussian noise proportional to the square root of the variance, where the variance in each image is obtained in ADU by dividing the model image by a gain set to 4.62 across all bands.  

\begin{figure}
\centering
\includegraphics[width=0.9\linewidth]{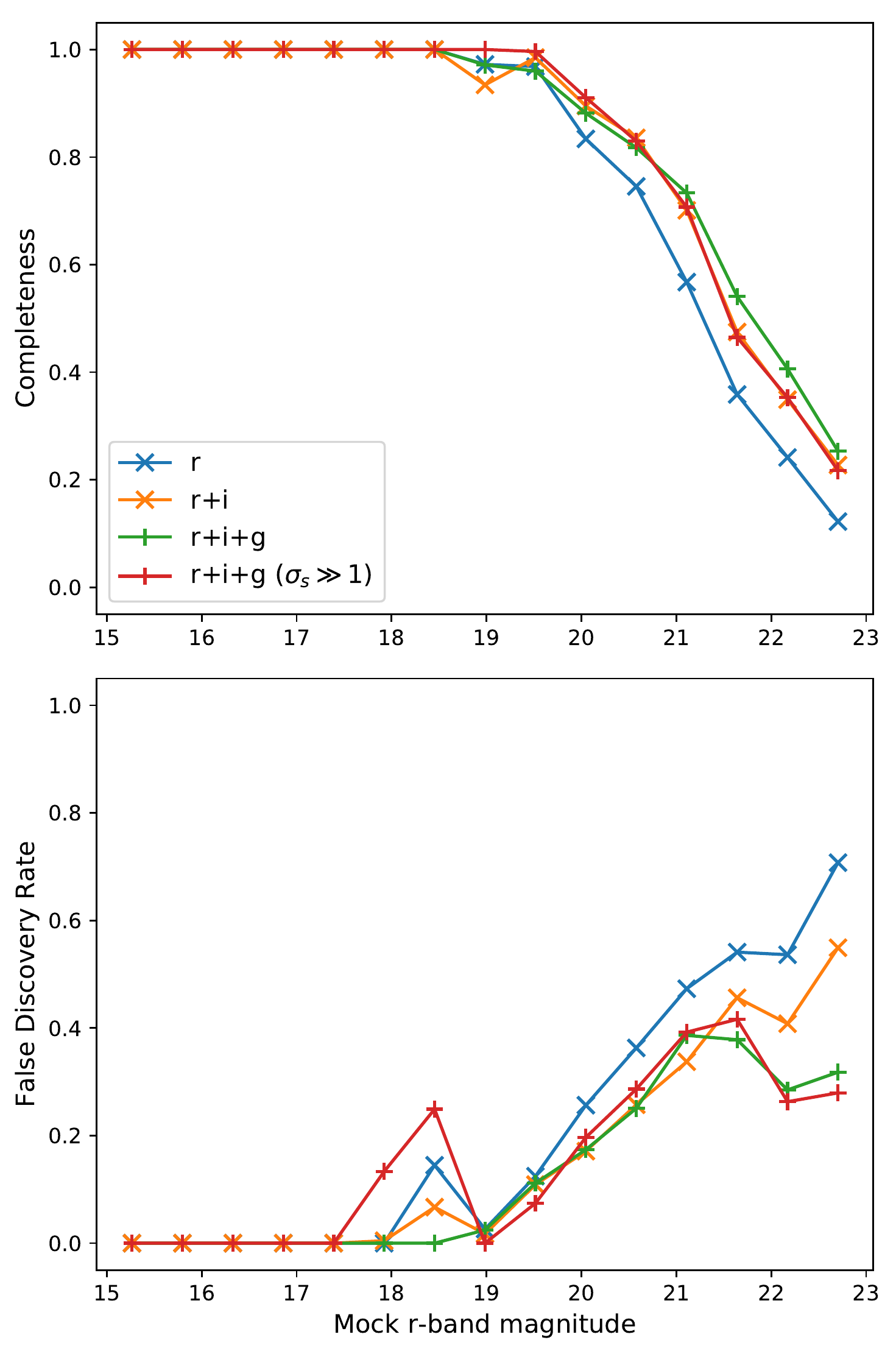}

\caption{Completeness (top) and false discovery rate (bottom) plots for mock PCAT runs on single-band (blue), two-band (orange), and three-band (green, red) realizations of a crowded mock catalog. The red plots come from a run on three-band data with effectively flat color priors ($\sigma_{r-i} = \sigma_{r-g} = 5$.)}
\label{fig:compare_bands_mock100}
\end{figure}

Figure \ref{fig:compare_bands_mock100} shows the completeness and false discovery rates for the catalog ensembles of one-, two-, and three-band PCAT runs. Because PCAT gives us an ensemble of catalogs, the completeness we report is an average over catalog samples. For each sample, we match a sample source with a true source if the positions of the two sources differ by less than 0.5 pixels and their $r$-band magnitudes are within 0.5 mag of each other. The choice of these matching criteria is explained in \S 6.2. We are primarily concerned with understanding the improvement that comes from adding multiple bands to the fit. 

We can estimate the improvement in completeness by calculating how the magnitude threshold for detecting a source changes with additional observations. Given a single isolated source with flux $f$ in a uniform background $B$, we wish to calculate $f'$, the flux that yields the same S/N by combining $n$ observations:
\begin{equation}
(S/N)^2 = \frac{f^2}{N_{eff}B+f} = \frac{nf'^2}{N_{eff}B+f'}
\end{equation}
Taking the approximation $N_{eff}B \gg f$, this reduces to
\begin{equation}
f' = \frac{f}{\sqrt{n}}
\end{equation}
For $f=250$ ADU ($r\approx 22.2$), we calculate $\Delta m = m'-m$ to be $0.37$ and $0.60$ for $n=2$ and $n=3$ observations, respectively. As seen in Figure \ref{fig:compare_bands_mock100}, the improvements in catalog completeness by adding $i$ and $g$-band to runs on mock data are consistent with these predictions. By comparing the two runs on three-band data (green and red lines), one can see that the color prior plays an important role in enhancing point-source detection past $r \sim 21$ and reducing false positives.     

\subsection{Astrometric flexibility}
In practice, astrometric calibration can be compromised by a number of instrumental and other systematics. To understand how astrometric miscalibration affects the multiband catalog ensemble, we take the same $100\times100$ pixel mock SDSS data and perturb what would otherwise be perfect astrometry. We consider a two-band case ($r$ and $i$), where $i$-band model source positions are artificially perturbed. 

\begin{figure}
\centering
\includegraphics[width=0.9\linewidth]{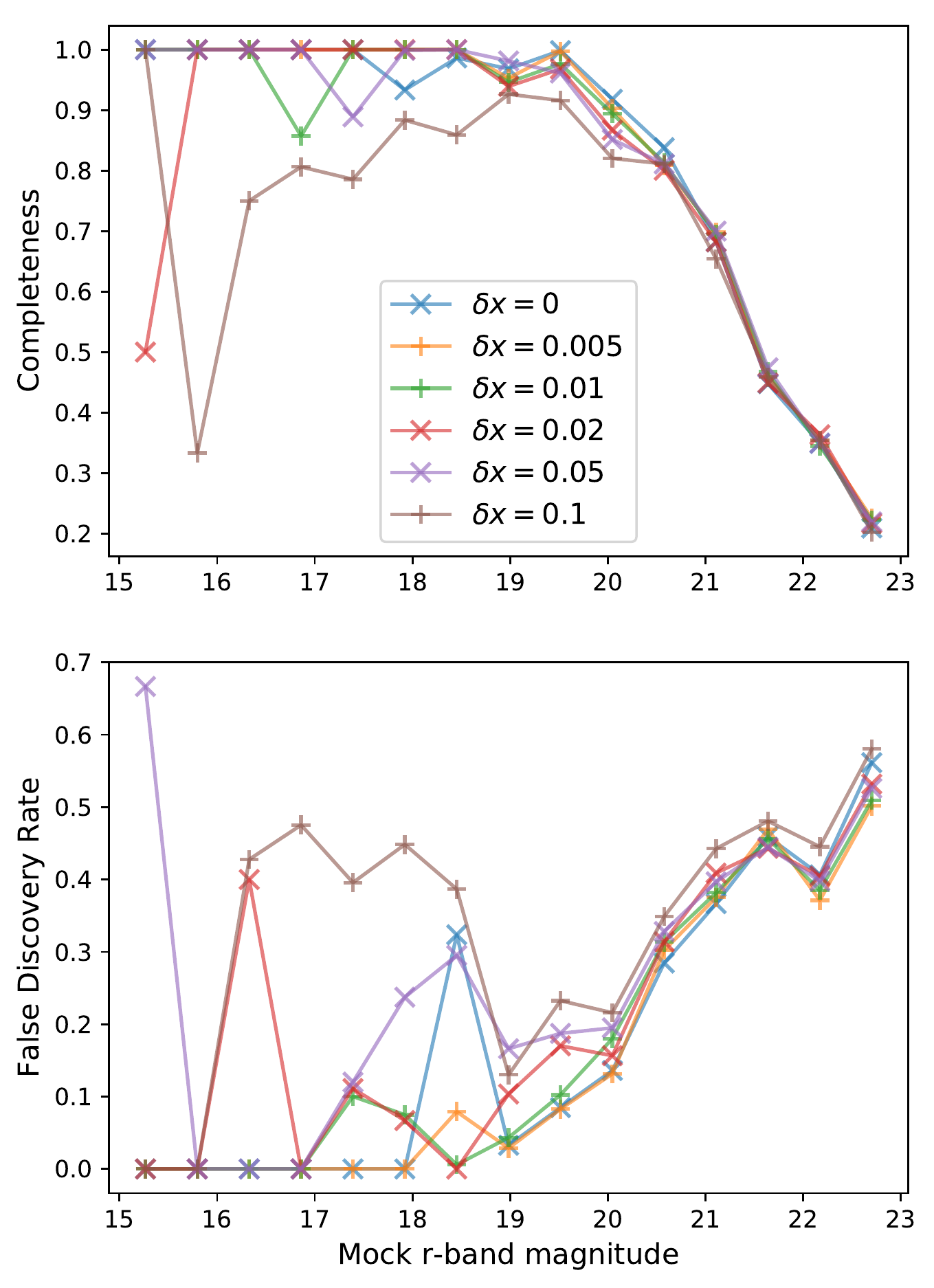}
\caption{Completeness (top) and false discovery rate (bottom) for an $r$- and $i$-band fit as a function of $r$-band magnitude of mock data. The different plots correspond to the results of runs with $i$-band astrometry perturbed by different values of $\delta x$. Larger perturbations cause source oversplitting, which corresponds to lower completenesses and higher false discovery rates.}
\label{fig:mock100_diff_astrometry}
\end{figure}

Figure \ref{fig:mock100_diff_astrometry} shows the catalog ensemble completeness and false discovery rates of multiband PCAT for cases with offsets $\delta x = 0.005, 0.01, 0.02, 0.05$, and 0.1 pixels. Subpixel perturbations of bright sources bias the delta log-likelihood of a sampling proposal. For a Gaussian problem, this can be approximated as $\Delta \log \mathcal{L} \sim (\delta x/2\sigma_x)^2$, where $\sigma_x$ is the uncertainty in a source's position. In these cases the likelihood may favor a scenario where bright sources (i.e. sources with small $\sigma_x$) are split into pairs of overlapping sources. Indeed, the brightest sources in our mock dataset get oversplit when the astrometry is perturbed at the level of $10^{-2}$ pixels, and the effect becomes more severe for larger miscalibrations. 

\section{Disentangling Blended Sources through Multiband Fitting}

In this section we examine in detail how multiband probabilistic cataloging may enhance source detection by directly disentangling blended sources. In particular, 
\begin{enumerate}
\item Can a multiband PCAT fit distinguish blended sources more reliably than a corresponding single-band analysis?  
\item If so, is the improvement primarily the result of higher S/N, or does color information play a role as well? 
\end{enumerate}

To test these ideas, we generate mock realizations of a scenario with two highly blended point sources. We test a number of configurations, varying the separation between sources along with their absolute and relative fluxes. The various configurations of these parameters are described in Table~\ref{tbl:two_src_params}. The colors of the two sources are $(r-i, g-r)_1 =(0.3, 0.5)$ and $(r-i, g-r)_2 = (0.1, 0.1)$, and were chosen because they both reside within a high-density region of the color prior and yet are spectrally distinct from one another. We set the backgrounds in each band to 179 ADU, the empirically derived $r$-band SDSS background level used in \S 3. We then generate a $30 \times 30$ pixel mock image for each scenario, with a fixed gain based on SDSS observations. Ten Gaussian noise realizations are generated and tested for each configuration.

\begin{figure*}[t]
\centering
\includegraphics[width=0.8\textwidth]{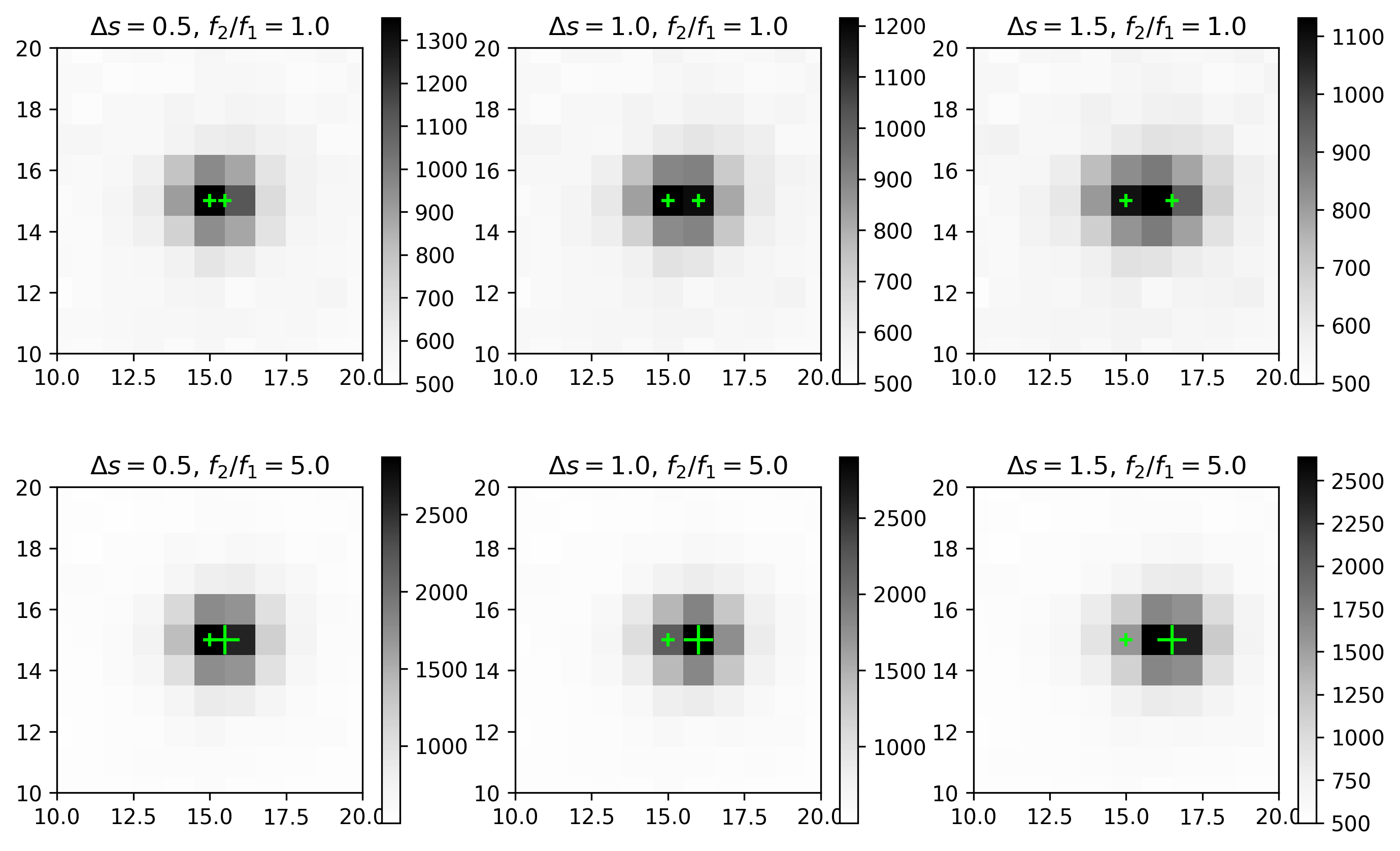}
\caption{Co-added mock images of two covariant sources at varying separations $\Delta s$, with $f_2/f_1=1$ (top) and $f_2/f_1 = 5$ (bottom). The green plus signs mark the true source positions within each image.}
\label{fig:2src_images}
\end{figure*}
To test how much improvement in the multiband case can be attributed to higher S/N or spectral information, we also generate equally weighted co-adds of $r$-, $i$-, and $g$-band images. These should serve as a reasonable reference point for multiband probabilistic cataloging; if color priors do inform the fit in a significant way, one would expect improvements reflected through the catalog ensemble. Figure \ref{fig:2src_images} shows co-added mock images of two covariant sources at different separations.

\begin{table}
\begin{center}
\begin{tabular}{cc}
Parameter & Values \\
\hline
\hline
Separation $s$ & 0.5, 0.6, 0.75, 1.0, 1.2, 1.5 \\
$f_2/f_1$ & 1.0, 2.0, 5.0 \\
$r$-band flux $f_1$ & 250 (4.3$\sigma$), 500 (8.3$\sigma$), 1000 (15.5$\sigma$) \\
No. of noise realizations & 10 \\
\end{tabular}
\end{center}
\caption{Values of parameters used in different two-source blending scenarios. The source separation $s$ has units of pixels, and fluxes are in ADU. These fluxes correspond to SDSS $r$-band magnitudes 22.2, 21.4 and 20.6, respectively. By convention, source 2 is defined to have equal or greater flux than source 1. The point source significances are calculated for a source with flux $f_1$ observed in one band.}
\label{tbl:two_src_params}
\end{table}

\subsection{Association Procedure}
Because the two sources are generated at small separations, the condensed catalog procedure from \cite{PORTILLO_17} is not feasible for making associations between catalog samples and true sources. Instead, a modified procedure is used to associate sample sources with the two mock sources. For consistency, we refer to the fainter of two sources as source 1 and the brighter as source 2. For each sample:

\begin{enumerate}
\item If the sample contains one source: 
\begin{itemize}
\item If $f_2/f_1 > 1$, then associate the sample with the brighter source, so long as its position does not deviate from that of the bright source by more than the separation of the two real sources. This is motivated by the fact that samples with a single source are likely blended versions of the two real sources, from which an association is stronger with the brighter source.
\item If $f_2/f_1 = 1$, associate each sample source with the closest mock source.  
\end{itemize}
\item If the sample contains two or more sources, determine the source closest to source 2 and make that association if it is within the separation offset $s$ of source 2. Then do the same with the closest remaining sample source for source 1.  
\end{enumerate}
In this procedure not all sample sources make associations, either because their positions deviate too far from the real sources or because they are in a catalog sample containing more than two sources and were not associated first. This procedure has not been optimized fully, but it should provide adequately robust source associations for the purposes of this study. 
\subsection{Source Prevalence Analysis}

\begin{figure*}[p]
\centering

\includegraphics[width=0.8\textwidth]{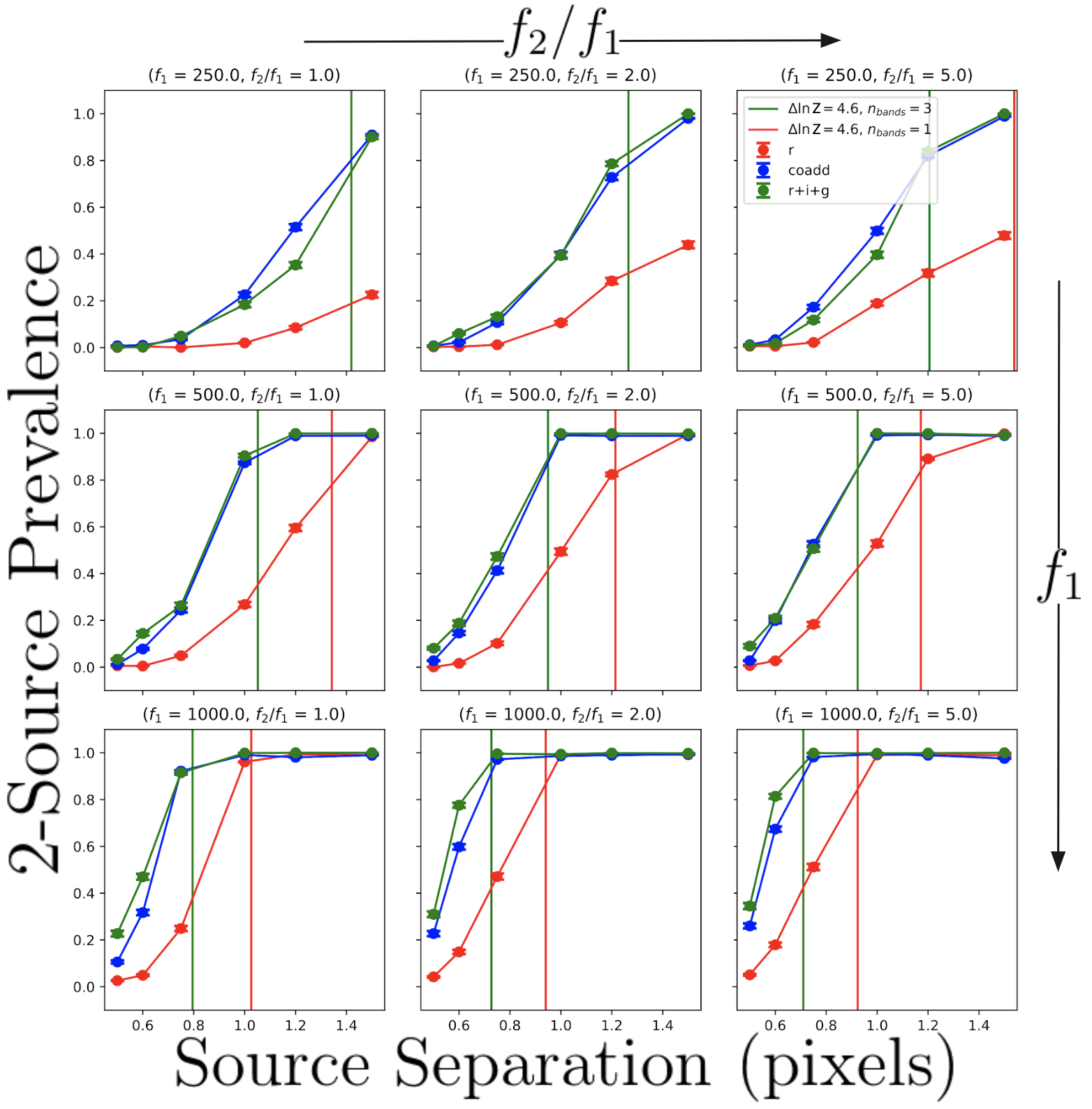}
\caption{Two-source prevalence as a function of source separation. Each point and corresponding error bars represent prevalence estimates from ten runs with different noise realizations. Results are shown for configurations with varying fluxes (units of ADU) and flux ratios between sources 1 and 2. Red plots denote runs on single $r$-band images, blue plots are from runs on co-added images, and green plots come from joint runs on $r$-, $i$-, and $g$-band images. The solid lines denote the estimated separations at which two-source descriptions of the data are favored over one-source descriptions by $\Delta \ln Z=4.6$. These separations are consistent with near-unity two-source prevalences from mock catalog ensembles.}
\label{fig:2star_prev}
\end{figure*}

To assess the performance of multiband PCAT in these various scenarios, the $n$-source prevalence $p_n$ (i.e. the fraction of catalog samples with source number $n$) is calculated for $n=1$, $n=2$, and $n\geq 3$. For a converged chain, the prevalence may be interpreted as the posterior probability for a catalog with $N_{star}=n$. While in \cite{PORTILLO_17} the term is used to refer to individual sources in a condensed catalog, the $n$-source prevalence here is label invariant. A similar source prevalence analysis can be found in \cite{JONES_15}, though our analysis does not explore the effect of relative background levels on the fit. In all runs, $p_{>2}$ was measured to be $\leq 0.01$. As such, two-source prevalences $p_2$ are shown in Figure \ref{fig:2star_prev} as a function of separation offset, from which $p_1 \approx 1-p_2$.\footnote{As a check, the same analysis was done on multiband datasets containing one source rather than two. The reason for this is that any oversplitting biases would propagate through all elements of analysis. After a burn in period of 100 thinned samples, the remaining samples for all chains returned one-source prevalences $p_1 \geq 0.99$, validating that PCAT does not split sources with fluxes across multiple bands.} 

Figure \ref{fig:2star_prev} compares the two-source prevalence as a function of separation between one-band ($r$), joint multiband ($r, i, g$), and co-add runs. As expected, the co-add and joint multiband PCAT runs detect two sources more often than corresponding single-band runs. 
As a check on PCAT's performance, we estimate the source separation that is needed to deblend two neighboring sources. The one- and two-source prevalences are related by the ratio of their Bayesian evidences:
\begin{equation}
    \frac{p_2}{p_1} = \frac{Z_2}{Z_1} = \frac{\pi(n=2)}{\pi(n=1)}\frac{\int \mathcal{L}(\theta_1, \theta_2) \pi(\theta_1, \theta_2) d\theta_1 d\theta_2}{\int \mathcal{L}(
    \theta_1) \pi(\theta_1) d\theta_1},
\end{equation}
where $\theta_{1,2}$ is the set of parameters describing the first or second star. For this derivation, we assume that the two-source likelihood is strongly peaked around the true parameters and is approximately an independent Gaussian for each parameter, ignoring possible covariance between parameters. We also approximate the prior as flat in the region where the likelihood is peaked, with an effective width equal to the reciprocal of the value of the prior at the true parameters $\theta^*_{1,2}$. Then, the evidence integral can be approximated as:
\begin{equation}
    \ln Z_2 \approx \ln \pi(n=2) + \ln \mathcal{L}(\theta_1^*, \theta_2^*) + \sum_{\theta \in \{\theta_1,\theta_2\}} \ln \frac{\sigma_\theta}{w_\theta}
\end{equation}
where $\sigma_\theta$ is the uncertainty on parameter $\theta$, estimated from the second derivative of the log-likelihood, and $w_\theta$ is the effective width of the prior. This approximation of the log evidence can be interpreted as the sum of the log prior for $n=2$, the maximum log-likelihood, and the log ratio of the uncertainty over the prior width for each parameter (these log ratios are always negative since the uncertainty must be smaller than the prior width). For the one-source likelihood, we assume that the likelihood is strongly peaked around parameters $\Tilde{\theta}$ corresponding to the total flux and center of flux, giving
\begin{equation}
    \ln Z_1 \approx \ln \pi(n=1)+ \ln \mathcal{L}(\Tilde{\theta}) + \sum_{\theta \in \theta_1} \ln \frac{\sigma_\theta}{w_\theta}.
\end{equation}
We approximate the delta log-likelihood between the true two-source parameters $(\theta_1^*, \theta_2^*)$ and the blended source parameters $\Tilde{\theta}$ by calculating
\begin{equation}
    \ln \mathcal{L}(\theta_1^*, \theta_2^*) - \ln \mathcal{L}(\Tilde{\theta}) \approx \sum_{b=1}^{n_{bands}}\sum_{l=1}^w\sum_{m=1}^H \frac{(\Delta\lambda_{lm}^b)^2}{2\lambda_{lm}^b}
\end{equation}
where $\Delta\lambda_{lm}^b$ is the difference between the two-source and blended source model images and $\lambda_{lm}^b$ is the two-source model image. These approximations neglect the possible covariance between the noise and the maximum likelihood two-source and one-source model parameters. The differences in the prior terms $\ln ( \sigma_\theta/w_\theta)$ contribute significantly to the difference in evidence: they sum to $\approx 22$, with the $x,y$ terms contributing the most at $\approx 6$ each. The flux and color terms contribute slightly less ($\approx 4$ and 3 each, respectively), while the prior on the number of sources contributes 0.5 for each additional degree of freedom (see \S A.2). Using $p_1 \approx 1-p_2$ and denoting $\Delta \ln Z = \ln Z_2 - \ln Z_1$,
\begin{equation}
    p_2 \approx \frac{e^{\Delta \ln Z}}{1 + e^{\Delta \ln Z}},
\end{equation}
so we report the separations where $\Delta \ln Z = 4.6$, corresponding to an expected $p_2 = 0.99$. These threshold source separations (also shown in Figure \ref{fig:2star_prev}) are consistent with near-unity two-source prevalences obtained using PCAT for both single-band and multiband runs.

The multiband runs done on $r$-, $i$-, and $g$-band data yield higher prevalences at small separations than corresponding runs on co-added data. This is evidence that imposing color priors can improve the fit of the resulting catalog ensemble. Note that the color priors used in these runs are fairly broad and uninformative. The degree of improvement one might expect from spectral constraints will roughly be a function of the fluxes and colors of the covariant sources, along with the color prior itself. A scenario where color priors can help the model favor two sources is when merging two covariant sources results in a brighter source with colors less favored by the priors. 

\begin{figure*}[p]
    \centering
    \includegraphics[width=\linewidth]{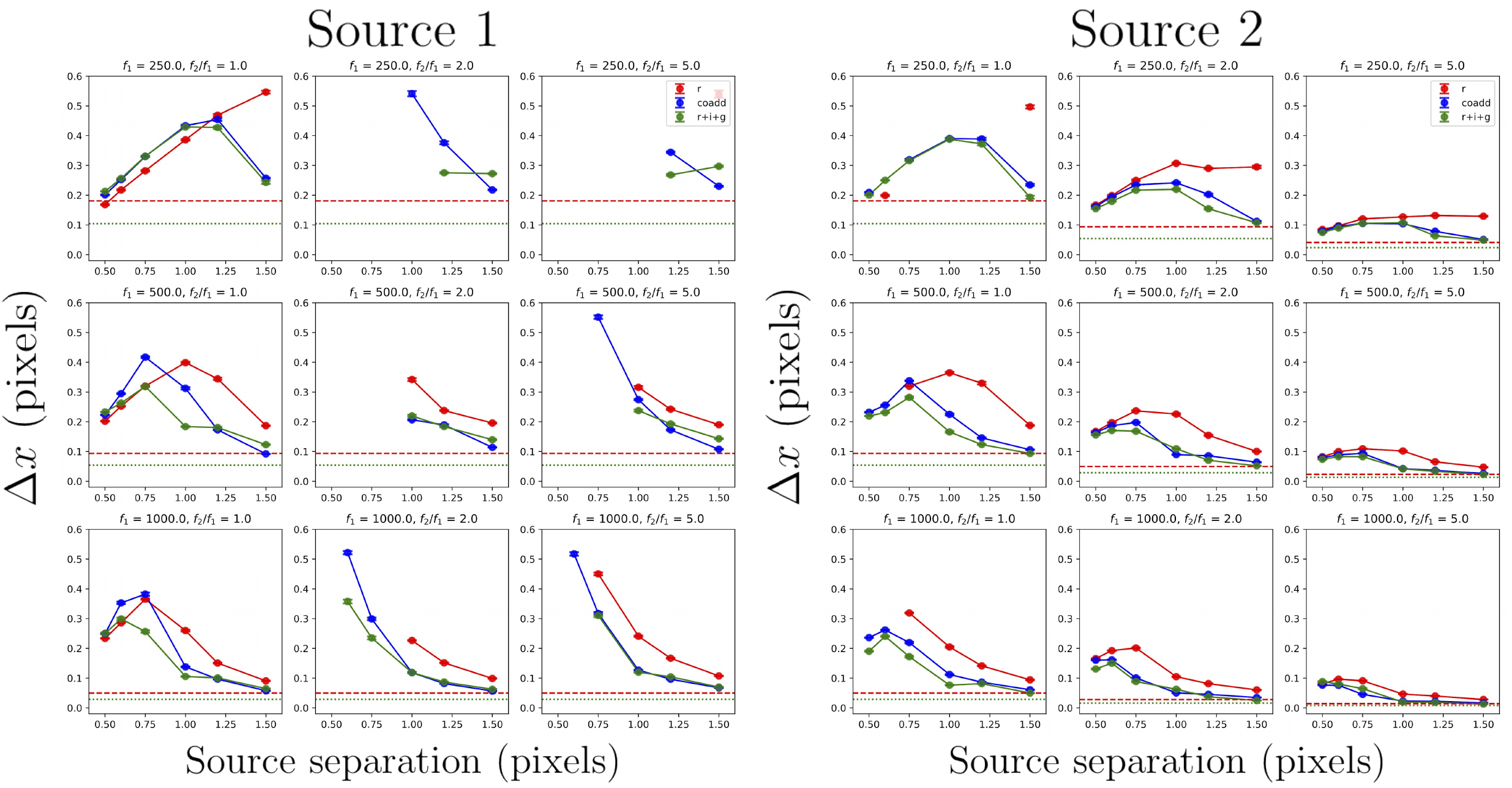}
    \caption{Mean position error for source 1 (left) and source 2 (right) as a function of source separation distance. Each point and corresponding error bars represent the error estimates from ten runs with different noise realizations. Results are shown for configurations with varying fluxes (units of ADU) and flux ratios between sources 1 and 2. Red plots denote runs on single $r$-band images, blue plots are from runs on co-added images, and green plots come from joint runs on $r$-, $i$-, and $g$-band images. The dashed horizontal lines show the expected error in the sparse field limit for a source observed in one band (red) and three bands (green). Model sources that did not meet the association criteria (see \S 4.1) to true mock sources are not included, which explains why a number of points from low-S/N runs are not plotted.}
    \label{fig:2src_pos}
    \centering

    \includegraphics[width=\linewidth]{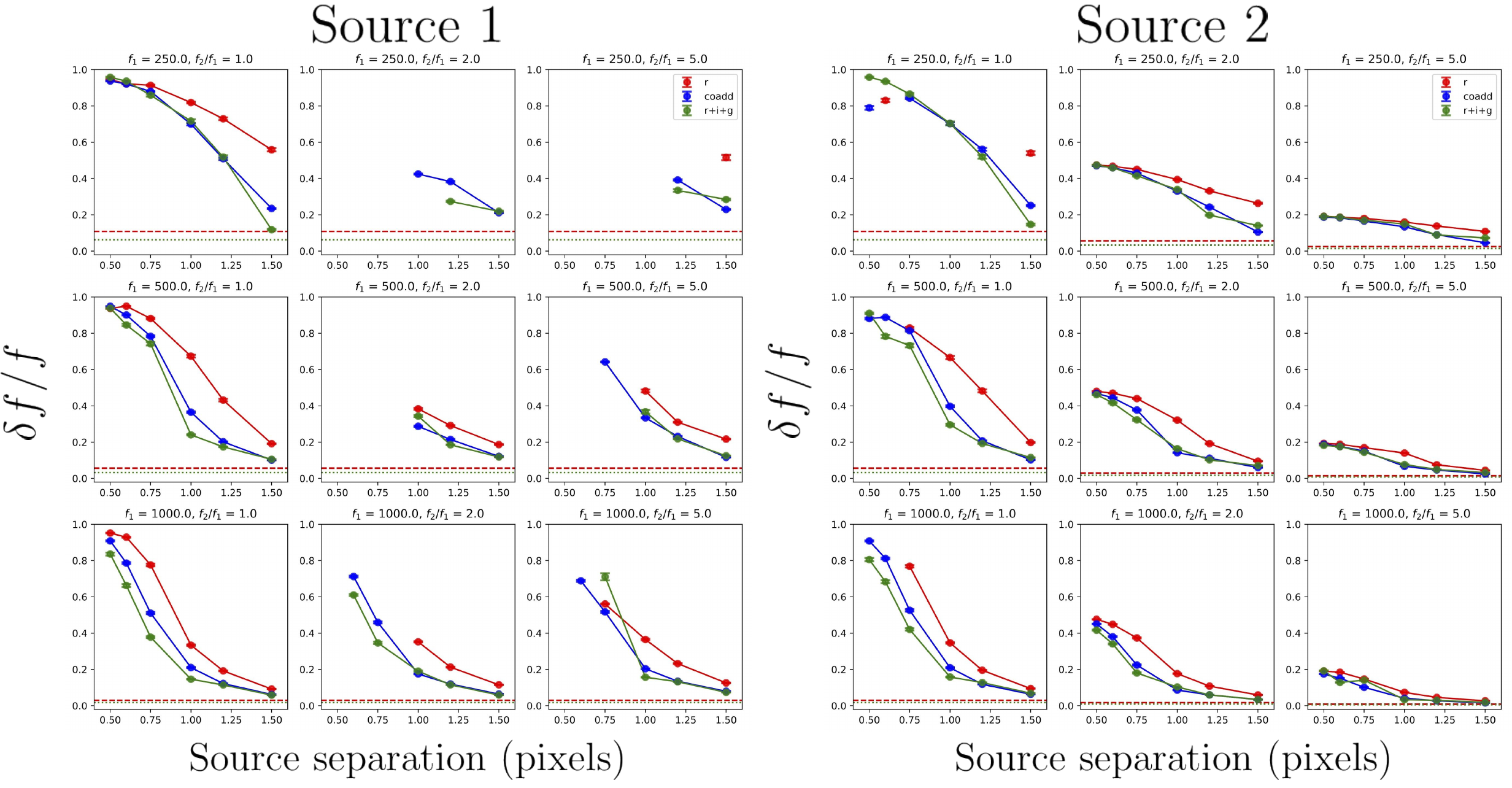}
    \caption{Same results as Figure \ref{fig:2src_pos} but for fractional flux errors.}
    \label{fig:2src_flux}
\end{figure*}

\subsection{Position and Flux Errors}

Using the source association procedure outlined in \S 4.1, one can calculate the mean error on position and flux for different two-source configurations. Figure \ref{fig:2src_pos} shows the mean absolute position errors for each source. The co-add and joint multiband runs recover more accurate source positions than corresponding single-band runs, which can be primarily attributed to higher S/N. Also plotted with dashed lines are the estimated position errors if one considers a source with no covariant neighbors and assumes that the likelihood dominates over the prior (see Appendix C of \cite{PORTILLO_17} for a derivation). As the separation between both sources increases, the position errors for single-band and multiband runs converge toward these lower bounds. This agreement demonstrates the robustness of probabilistic cataloging, especially when one considers that the sources are still highly covariant at these separations.    

Similar improvements can be seen in Figure \ref{fig:2src_flux}, which shows the fractional flux error for both sources as a function of separation. Errors in flux may come from one sample source absorbing the flux from two highly covariant mock sources, or from two covariant sample sources with incorrectly distributed fluxes.

\section{Source Coordinate Transformations across Observations}

As demonstrated in \S 3, the coordinate transformations from one band to another must be accurate to subpixel levels in order for PCAT to successfully forward-model source emission in all bands simultaneously. While standard astrometry packages can achieve this level of accuracy when detector distortion coefficients are available, they are often slow, employing iterative methods and the inversion of high-order polynomials. The computational expense of these standard transformations is negligible when they are only executed a handful of times. PCAT, however, requires $\sim 10^6$ model perturbations in its latest implementation to converge on the posterior, transforming the coordinates of $\sim 10^3$ sources the same number of times along the way. In this section, we demonstrate that a combination of pixel-to-sky mappings from SDSS and per-pixel linear interpolation is sufficiently accurate and yields a significant computational speedup compared to standard methods. While we use SDSS as an example for this work, we note that the information required to do these transformations should be provided by most modern telescope surveys. 

\subsection{Sloan Digital Sky Survey}
SDSS is an imaging and spectroscopic survey designed to probe large-scale structure \citep{SDSS}. Located at the Apache Point Observatory in New Mexico, the SDSS telescope is a 2.5m, $f/5$ modified Ritchey-Chr\'etien wide-field altitude-azimuth telescope. Since it began collecting data in 2000, SDSS has covered over one-third of the sky, with over 500 million objects detected photometrically. Of those, nearly three million have observed spectroscopically with either the APOGEE spectrograph, used to study galactic structure in the Milky Way \citep{APOGEE}, or the BOSS spectrograph, which gathers galaxy redshift information to probe cosmic expansion through baryon acoustic oscillations \citep{EISENSTEIN}. 
 
\subsubsection{SDSS Photometric Observation}
The SDSS camera consists of 30 $2048\times 2048$ pixel SITe/Tektronix CCDs, with a pixel size of 24 $\mu$m, which corresponds to 0.396" on the sky \citep{SDSS}. Each of SDSS's five optical bands, namely, \textit{u,g,r,i,z}, corresponds to a row of CCDs on the camera.\footnote{\url{http://www.sdss.org/instruments/camera/}} The telescope uses a drift scanning technique, in which the camera takes continuous observations as the telescope moves in great circles along the sky. The camera reads data from each CCD at a rate equal to the drift rate, nominally the sidereal rate ($\sim 15^{\circ} \text{hr}^{-1}$). The use of drift scanning minimizes the observational overhead one would typically incur from easing out the detector after each observation.  Each observing period is referred to as a \textit{run} lasting up to 11 hr. Each column, or \textit{camcol} of the camera images a strip of sky with a width of 2048 pixels. The strip from each CCD is continuous and arbitrarily divided into frames of 1361 pixels augmented by a 128-pixel overlap, making $1489\times 2048$ pixel frames the image size for processing. The five frames aligned by position on the sky are then called a \textit{field}. More information on the SDSS photometric observational setup can be found in \cite{GUNN}.

\subsubsection{SDSS Astrometric Calibration}
In order to map pixel coordinates to celestial coordinates, the SDSS processing pipeline divides the total drift scanned data into frames and proceeds with astrometric calibration within each frame. For each frame, the pipeline first makes corrections to the original pixel coordinates $(x,y)$ using a smooth cubic fit. 

For color $< \text{(color)}_0$:

\begin{align}
x' &= x + g_0 + g_1y + g_2y^2 + g_3y^3 + p_x (\text{color}) \\ y' &= y + h_0 + h_1y + h_2y^2 + h_3y^3 + p_y (\text{color})
\label{eq:linear_color}
\end{align}
For $\text{color} \geq (\text{color})_0$:

\begin{eqnarray}
\label{eq:constant_color}
x' = x + g_0 + g_1y + g_2y^2 + g_3y^3 + q_x \\
y' = y + h_0 + h_1y + h_2y^2 + h_3y^3 + q_y
\end{eqnarray} 
The coefficients $\lbrace g_i, h_i\rbrace_{i=1}^3$ and $\lbrace p_j, q_j\rbrace_{j\in \lbrace x,y\rbrace}$ are fit for each frame.

These transformations from \cite{PIER} correct for optical distortions that are a function of column only. This is the direction of the drift scan and of differential chromatic refraction (DCR). The variation from optical distortions is typically of order $\sim 0.2$ pixels, so the cubic fit accounts for that properly. DCR is modeled either as a linear function of color ($r-i$ for $r$, $i$, and $z$; $g-r$ for $g$; and $u-g$ for $u$) or as a constant depending on the observing band and the color of the observed source. Once the pixel coordinates are corrected, pixels are mapped to Catalog Mean Place (CMP) celestial coordinates $(\mu, \nu)$, the angular offsets along and perpendicular to the idealized great circle being scanned along. This is done through an affine transformation:

\begin{eqnarray}
\mu_{CMP} = a + bx' + cy'\\
\nu_{CMP} = d + ex' + fy'
\end{eqnarray}

The mappings of these transformations are contained for each frame within SDSS \texttt{asTrans} files.\footnote{For a more detailed description of these files, see \url{data.sdss.org/datamodel/files/PHOTO_REDUX/RERUN/RUN/astrom/asTrans.html}}  

\subsection{Linearization of astrometric transformations}

Consider the pixel transformation between bands $a$ and $b$. To transform $ \mathbb{R}^2: (x, y)_a \rightarrow (x, y)_b$, we separate $(x, y)_a$ into their integer and noninteger parts $(x_0 + dx, y_0 + dy)_a$ by truncation and use the following:
\begin{equation}\label{x_lin_interp}
x_b' = x_b(x_{a,0},y_{a,0}) + \Big(\frac{dx_b}{dx_a}\Big)dx_a + \Big(\frac{dx_b}{dy_a}\Big)dy_a
\end{equation}
\begin{equation}\label{y_lin_interp}
y_b' = y_b(x_{a,0},y_{a,0}) + \Big(\frac{dy_b}{dx_a}\Big)dx_a + \Big(\frac{dy_b}{dy_a}\Big)dy_a
\end{equation}
In the equations above, $x_b(x_{a,0},y_{a,0})$ and $y_b(x_{a,0},y_{a,0})$ are the integer-value transformed pixels taken directly from the asTrans files. The partial derivatives used in Equations \eqref{x_lin_interp} and \eqref{y_lin_interp} are approximated for each pixel with a first difference:\footnote{The denominator in these expressions have implicit units of pixels, since we are calculating the shift over a range of $(q_0+1) - (q_0-1) = 2$, where $q_0$ is a general coordinate.}
\begin{eqnarray}
\frac{dx_b}{dx_a}
 = \frac{x_b(x_{a,0} + 1, y_{a,0}) - x_b(x_{a,0} - 1, y_{a,0})}{2}\\
\frac{dx_b}{dy_a} = \frac{x_b(x_{a,0}, y_{a,0} + 1) - x_b(x_{a,0}, y_{a,0} - 1)}{2}
\end{eqnarray}
where $dx_b/dx_a$ and $dx_b/dy_a$ are evaluated at $(x,y)=(x_{a,0},y_{a,0})$.

This interpolation scheme allows us to map subpixel values by leveraging the information from local astrometric variation. In practice, these quantities are precomputed -- six arrays specify $[x_0]_b$, $[y_0]_b$, $\left[\frac{dx_b}{dx_a}\right]$, $\left[\frac{dx_b}{dy_a}\right]$, $\left[\frac{dy_b}{dx_a}\right]$, and $\left[\frac{dy_b}{dy_a}\right]$ for pixels in the desired region. In this scheme a pixel transformation is computed by six array lookups followed by multiplication and addition operations, rather than the typical trigonometric transformations used in WCS evaluations.

\subsection{Performance}

\begin{figure}
\centering
\includegraphics[width=\linewidth]{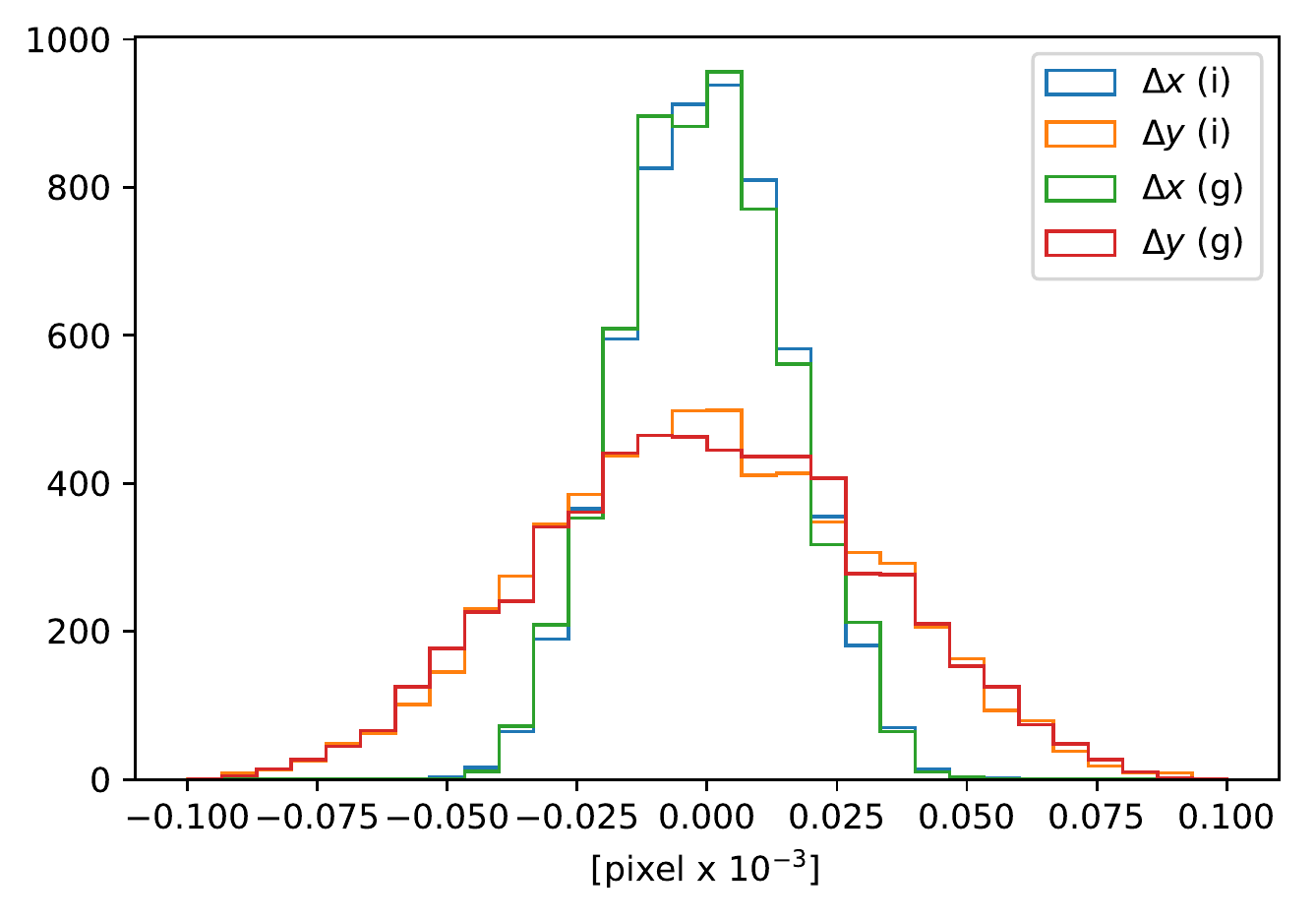}
\caption{Difference between source coordinates obtained by transforming from sky coordinates to the band of interest and those obtained by first transforming to $r$-band and using linearized asTrans coordinate transformations to get positions in band of interest. Histogram has units of millipixels. }
\label{fig:asTrans_resid}
\end{figure}

Figure~\ref{fig:asTrans_resid} shows the residual errors of linearized transformations from $r$ to $i$ and $g$, compared to the direct mappings from sky to $i$-band or $g$-band. The errors in $x$ and $y$ are centered about zero and below $10^{-4}$ pixels. As part of validation we also ran ``round trip" tests, using separate linearized transformations to take a set of coordinates from $r\to g$ and then back from $g\to r$. The residuals in positions from that test were below $10^{-5}$ pixels and centered about zero, likely the result of numerical round-off errors. While there may be larger systematic errors that affect the astrometry, these tests validate our method as consistent and unbiased.

\begin{figure}
\centering
\includegraphics[width=\linewidth]{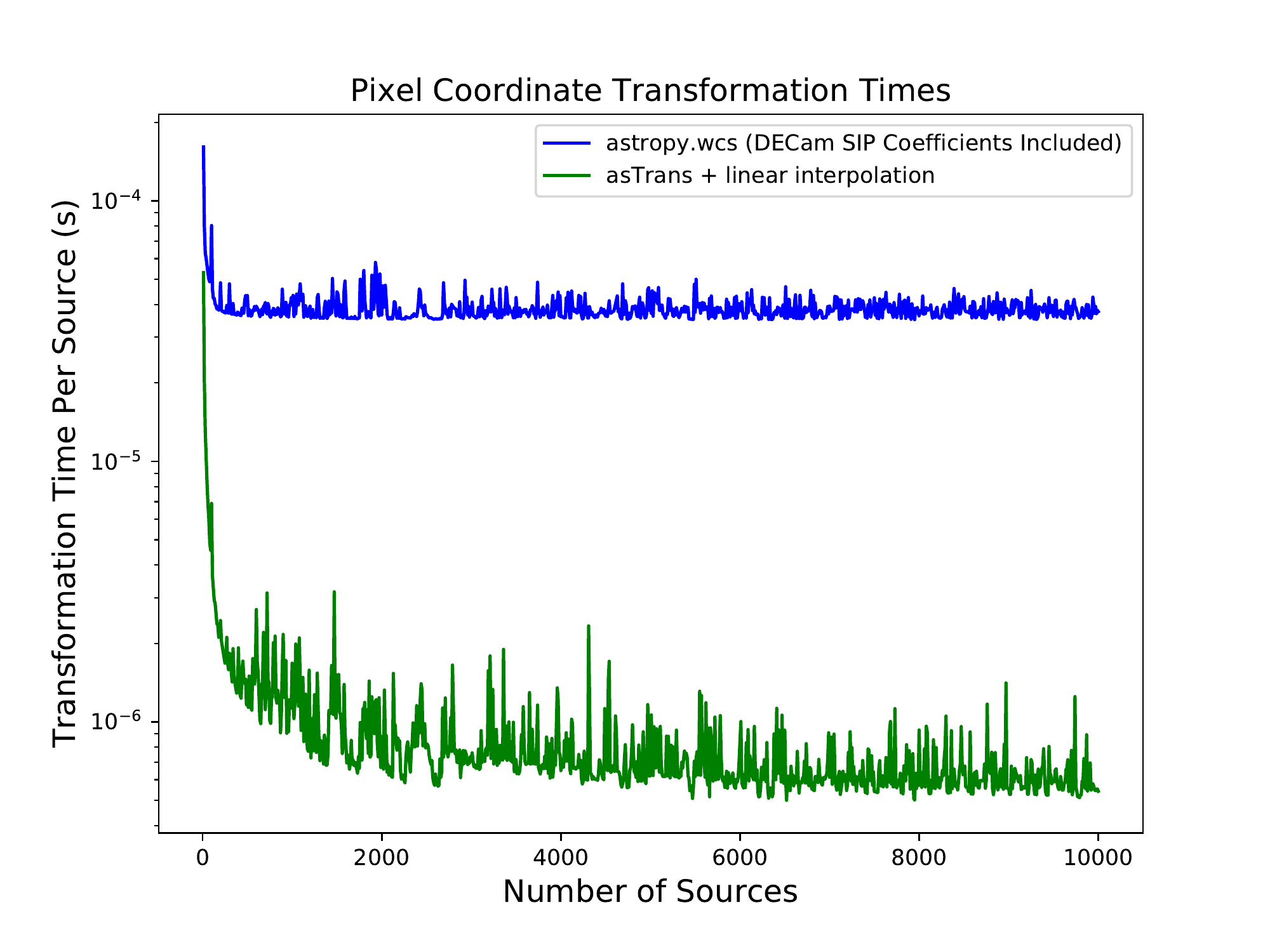}
\caption{Plot of the transformation time per source as a function of the number of sources transformed. The blue line displays results using the standard \texttt{astropy.wcs} Python coordinate transformation package, while the green line shows results for our method of asTrans mappings with linear interpolation.}
\label{fig:pix_trasf_times_wcs_astrans_lin}
\end{figure}

\begin{figure}
\centering
\includegraphics[width=\linewidth]{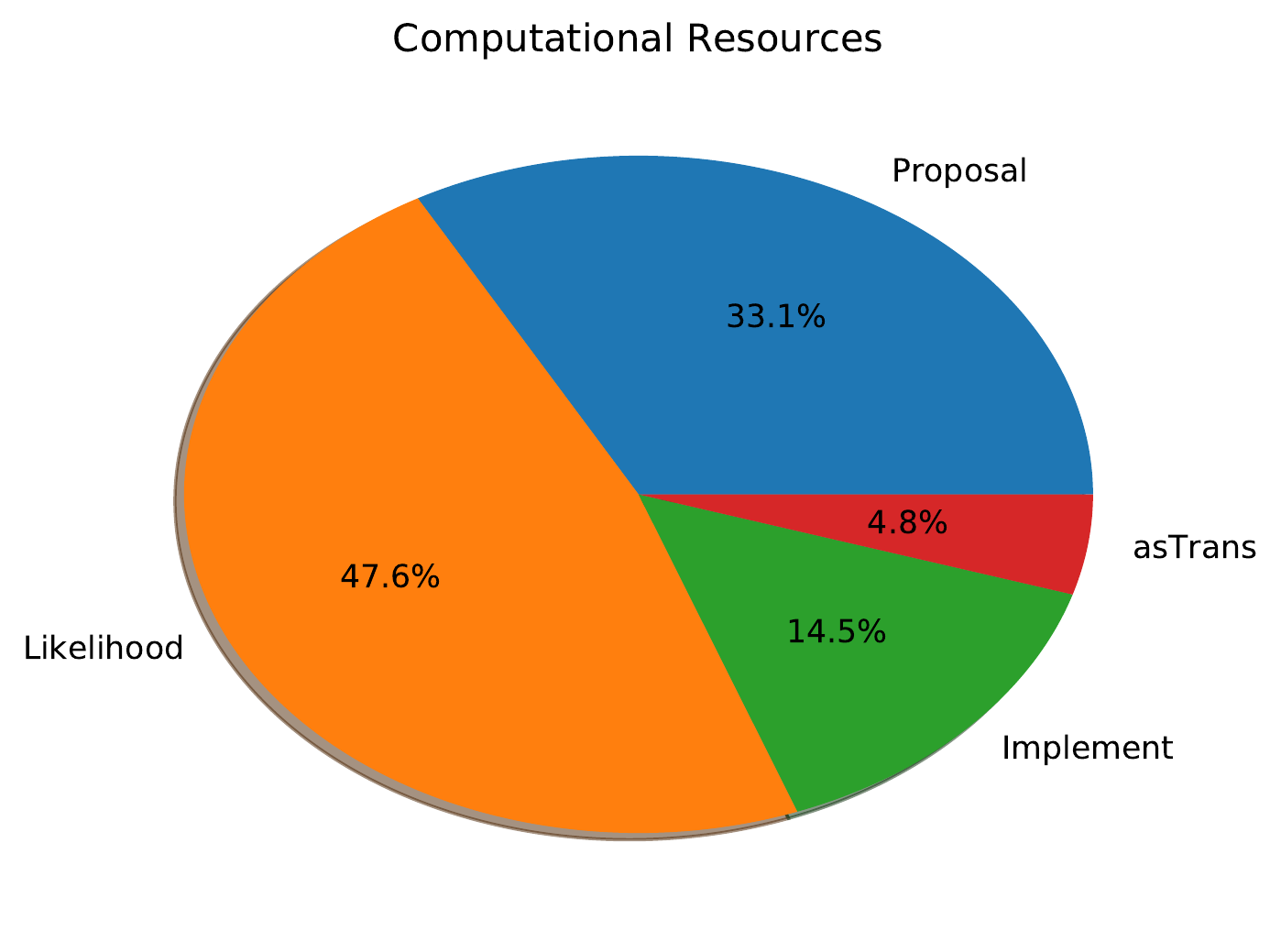}
\caption{Pie chart showing the computational resources used in a multiband PCAT run on three-band SDSS data. Computational resources are dominated by likelihood evaluation (orange), proposal generation (blue), proposal implementation (green), and, lastly, astrometric transformations (red).}
\label{fig:comp_resources}
\end{figure}

Figure~\ref{fig:pix_trasf_times_wcs_astrans_lin} shows a comparison of coordinate transformation times per source between \texttt{astropy.wcs}, a commonly used routine, and our linearized asTrans mappings. As the number of simultaneously perturbed sources increases, linearized transformations quickly outperform \texttt{astropy.wcs}; for $N_{src}=10^3$, there is a 40-fold speed improvement, and this grows to a factor of 80 by $N_{src}=10^4$. To emphasize the importance of having fast coordinate transformations, Figure \ref{fig:comp_resources} shows a breakdown of computational resources used in a run where $\sim 10^3$ sources were inferred. While the linearized coordinate transformations compose $\sim 5\%$ of all computations, \texttt{astropy.wcs} would use $\sim 200\%$ of the \emph{total} computational resources used by PCAT in its current form.

\section{Application to SDSS Data}
\subsection{Observations}
In this section, we present the results of multiband probabilistic cataloging applied to SDSS imaging of the globular cluster Messier 2 (M2). M2 is located 11.5 kpc away and contains about 150,000 stars with an overall stellar spectral type of F4 (\cite{harris_96}, 2010 edition). Our tests focus on the same $100\times 100$ pixel $r$-band cutout used in \cite{PORTILLO_17}, with the addition of SDSS $i$- and $g$-band data. The cutout comes from run 2583, camcol 2, field 136, with pixel coordinate $(x_0,y_0) = (310, 630)$ in the frame file defining the lower left corner. The center of the cutout in celestial coordinates is given by R.A. = $21^h 33^m 20.5^s$, decl. = $-0^{\circ} 48' 26.5"$. This field serves as a good test case because it is extremely crowded. It is crowded, in fact, that the SDSS photometric pipeline \texttt{Photo} originally timed out while trying to catalog the region. \cite{AN_2008} produce a catalog of the same region using DAOPHOT on SDSS \emph{ugriz} data. \cite{HST_CATALOG} provide a catalog of M2 derived from the HST Advanced Camera for Surveys (ACS). Both catalogs serve as important points of comparison. The DAOPHOT catalog of SDSS data represents results from a traditional approach to crowded field photometry. HST detects sources as faint as $r\sim 30$ because of superior sensitivity and an angular resolution $\sim 20$ times better than the Sloan telescope, meaning that its catalog can be treated as our reference ``truth" catalog. Along with smaller pixel sizes, observations from HST are not affected by atmospheric seeing. The region we use has been observed by HST in the ACS F606W and F814W bands. We compare with F606W, which covers a similar bandpass to the SDSS $r$-band.   

The bias and gain of each band are obtained from the opECalib CCD electronics calibration file\footnote{Description here: \url{https://data.sdss.org/datamodel/files}}. Next, the asTrans coordinate mappings described in \S 5 are obtained for $r \to i$ and $r \to g$. First, we generate mappings in which the mean color term is taken to be zero. We then incorporate the DCR correction by extracting coefficients $p_x$ and $p_y$ of Equation \eqref{eq:linear_color} from the asTrans files. The threshold colors $(g-r)_0 = 1.5$ and $(r-i)_0 \gg 1$ are sufficiently large such that we assume that all sample positions can be corrected with Equation \eqref{eq:linear_color}. DCR corrections are computed for each source at every step in the Markov chain and are typically at the level of $10^{-2}$ pixels. 

In this work, a fixed PSF template is used for each band. This assumes that the PSF does not vary over the region of interest. Because the region we analyze is a $100\times 100$ pixel region of M2 covering roughly $0.43 \text{ arcmin}^2$, such an assumption is reasonable. The PSF templates in $r$, $i$, and $g$ are extracted from SDSS \texttt{psField} files and are upsampled using a Lanczos kernel. As shown in \cite{PORTILLO_17}, the quality of the PSF has a dramatic effect on the quality of the fit. This is important because the PSF fitting done by SDSS is unvalidated for crowded fields like the one we examine. While the SDSS pipeline does a good job fitting the $r$- and $i$-band PSFs, we find the $g$-band PSF is notably broader, and inspection of residuals confirms that it is poorly estimated. To address this, we use the crowded field photometry code \texttt{crowdsource} \citep{Schlafly_2018} to refit the $g$-band PSF. We fit the PSF using a larger $600\times 600$ pixel region centered on our image for more constraining power.

\subsection{Catalog Ensemble}
Catalog ensembles for two multiband runs are obtained, one using $r$- + $i$-band and one using $r$- + $g$-band data. To ensure that samples in the catalog ensemble are drawn from the posterior distribution, we discard the first 1500 of 3000 thinned samples of the chain as burn-in. In the case of $r + i$, PCAT infers $1380 \pm 10$ sources, while for $r + g$ $1233 \pm 5$ sources are inferred. The corresponding single-band catalog ensemble contains $\approx 1100$ sources in the same region. DAOPHOT only detects 356 sources in this region, whereas HST detects 6051 sources down to F606W$\sim 30$, 1428 of which have F606W $< 23$.

In our first results, we observed oversplitting of some bright sources. We attribute this to astrometric miscalibration across bands -- \cite{PIER} report relative astrometric accuracy of $\sim 25-35$ mas between $r$-band and $u, g, i, z$. As discussed in \S 3.2, astrometric miscalibration at the level of $10^{-2}$ pixels is significant enough to affect sources at the bright end. To correct for this miscalibration, we take the brightest sources down to $r=19$ and, for each source, absorb any neighboring sources within 1 pixel. Upon inspection, the oversplit adjacent sources are almost always low-significance sources. Once these are recombined, the corresponding fluxes and positions are recalculated. While we were able to reduce oversplitting in two-band runs with this correction, astrometric miscalibration had a more severe impact on the $r+i+g$ three-band joint fit, the results of which we do not include in this work.    

\begin{figure}
    \centering
    \includegraphics[width=\linewidth]{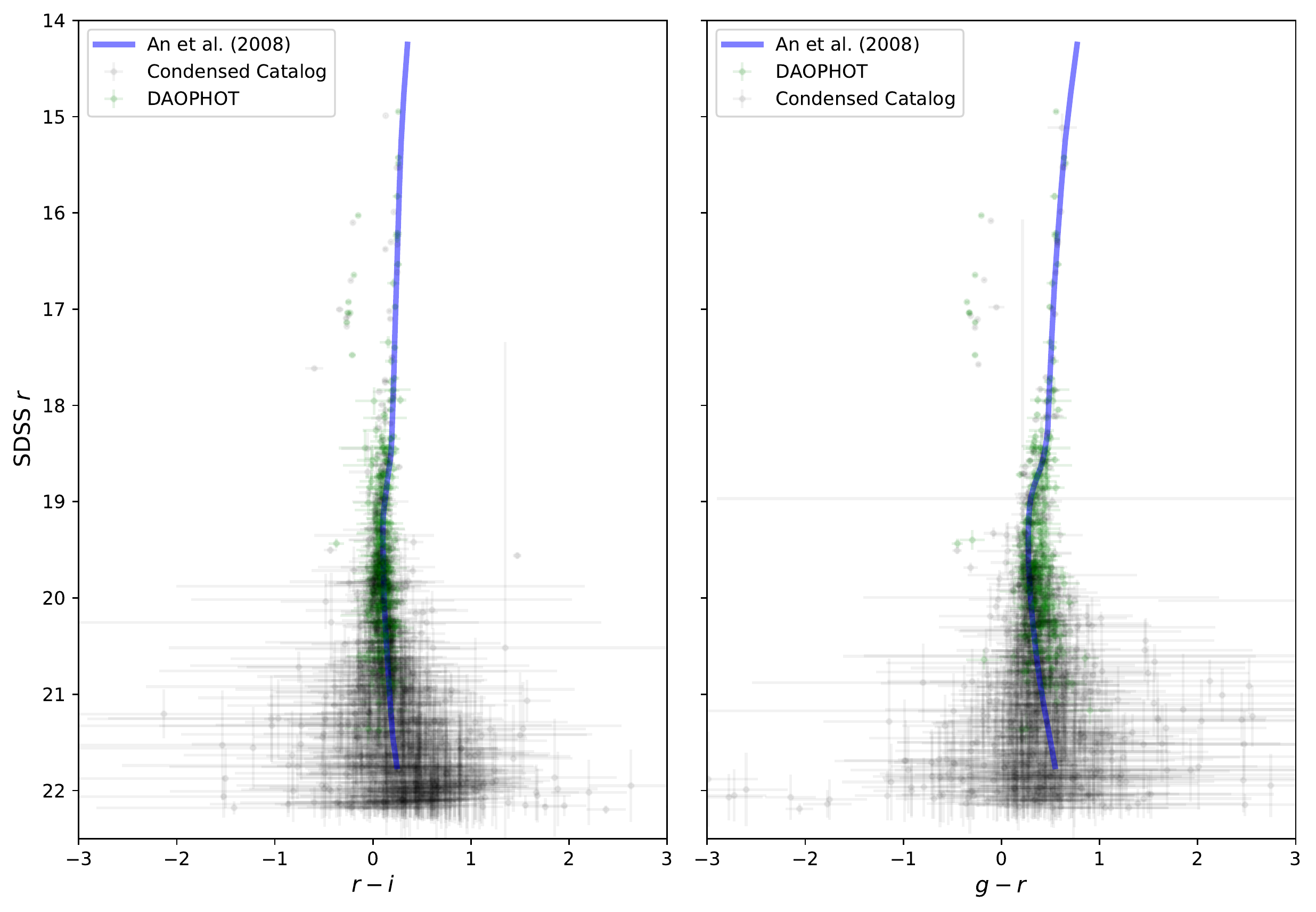}
    \caption{Color-magnitude diagrams for $r$- + $i$-band (left) and $r$- +$g$-band (right) condensed catalogs. Error bars on the condensed catalog come from marginalizing over catalog ensemble samples. Green points mark DAOPHOT catalog sources. Plotted in blue is the fiducial sequence for the full cluster M2 obtained by \cite{AN_2008}.}
    \label{fig:color_magnitude_2583}
    \includegraphics[width=0.47\linewidth]{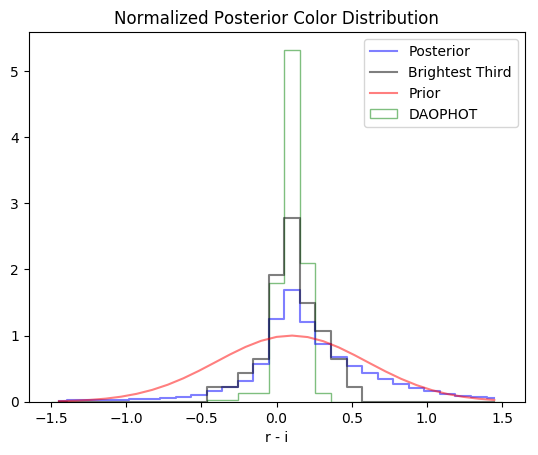}
    \includegraphics[width=0.48\linewidth]{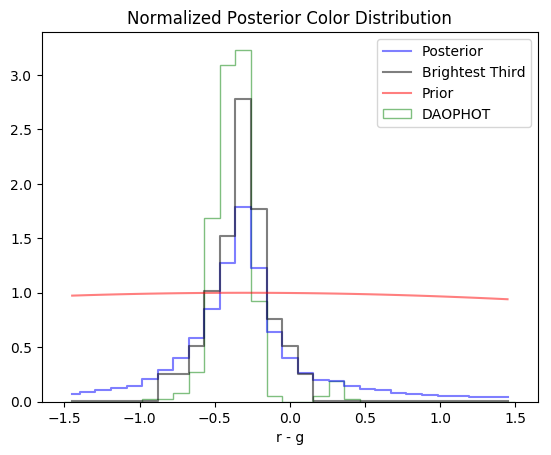}
    \caption{Posterior color histograms for $r+i$ (left) and $r+g$ (right) catalog ensembles. DAOPHOT color posteriors are shown in green, while catalog ensemble posteriors are shown in blue. Color posteriors from the brightest third of sources in each sample (sources with $r \lesssim 20.7$) are shown in black. These brighter sources should be similar to sources detected using DAOPHOT, and this similarity is reflected through similar posteriors. The color priors used in each fit (red) are relatively broad, though $\pi(r-i)$ is narrower than $\pi(g-r)$. }
    \label{fig:color_posterior}
\end{figure}

Color-magnitude diagrams for the two-band runs are shown in Figure \ref{fig:color_magnitude_2583}. To obtain uncertainties for individual sources, we reduce the last 300 catalog samples to a ``condensed catalog," which naturally marginalizes over the posterior catalog ensemble while providing a labeling that enumerates sources across different samples like a traditional catalog does. The process for producing a condensed catalog is outlined in \S 5 of \cite{PORTILLO_17}. We institute a modest cut on the condensed catalog, removing sources present in less than 10\% of samples. These are spurious sources that do not improve the catalog. Bright sources with large error bars are likely oversplit sources that are not converged. Also plotted are DAOPHOT catalog sources (green) and the fiducial sequence from \cite{AN_2008} for the full globular cluster. 

Posterior color distributions for each two-band run are displayed in Figure \ref{fig:color_posterior}. While a Gaussian prior $\pi(r-i) \sim \mathcal{N}(0.25, 1.0)$ is used, the posterior skews toward positive colors. Fainter sources in the catalog ensemble are less constrained by the data, which helps explains why brighter catalog sources have a posterior (black) that is narrower and more consistent with that of DAOPHOT catalog sources (green). 

\subsection{Completeness/False Discovery Rate}

To compare our catalog ensemble with the HST catalog, we associate sample sources with HST sources using two criteria: if a source is within 0.5 pixels of the HST position and has an $r$ magnitude within 0.5 mag of the F606W magnitude, it is considered a match. These matching criteria are motivated by the fact that F606W is a wider band than $r$, and SDSS sources will be biased brighter owing to faint neighboring sources HST can resolve but SDSS cannot. In the context of source detection, the estimates of completeness and false discovery are not significantly affected by these criteria. If one assumes that the positions of sources in our region are generated from a spatial Poisson process (i.e., uniformly throughout the region), the expected fraction of spurious associations with the HST catalog given our criteria is $\sim 3\%$ at $r=22$ and drops to $<1\%$ for $r<20$.     

\begin{figure*}[p]
    \centering
    \includegraphics[width=0.48\linewidth]{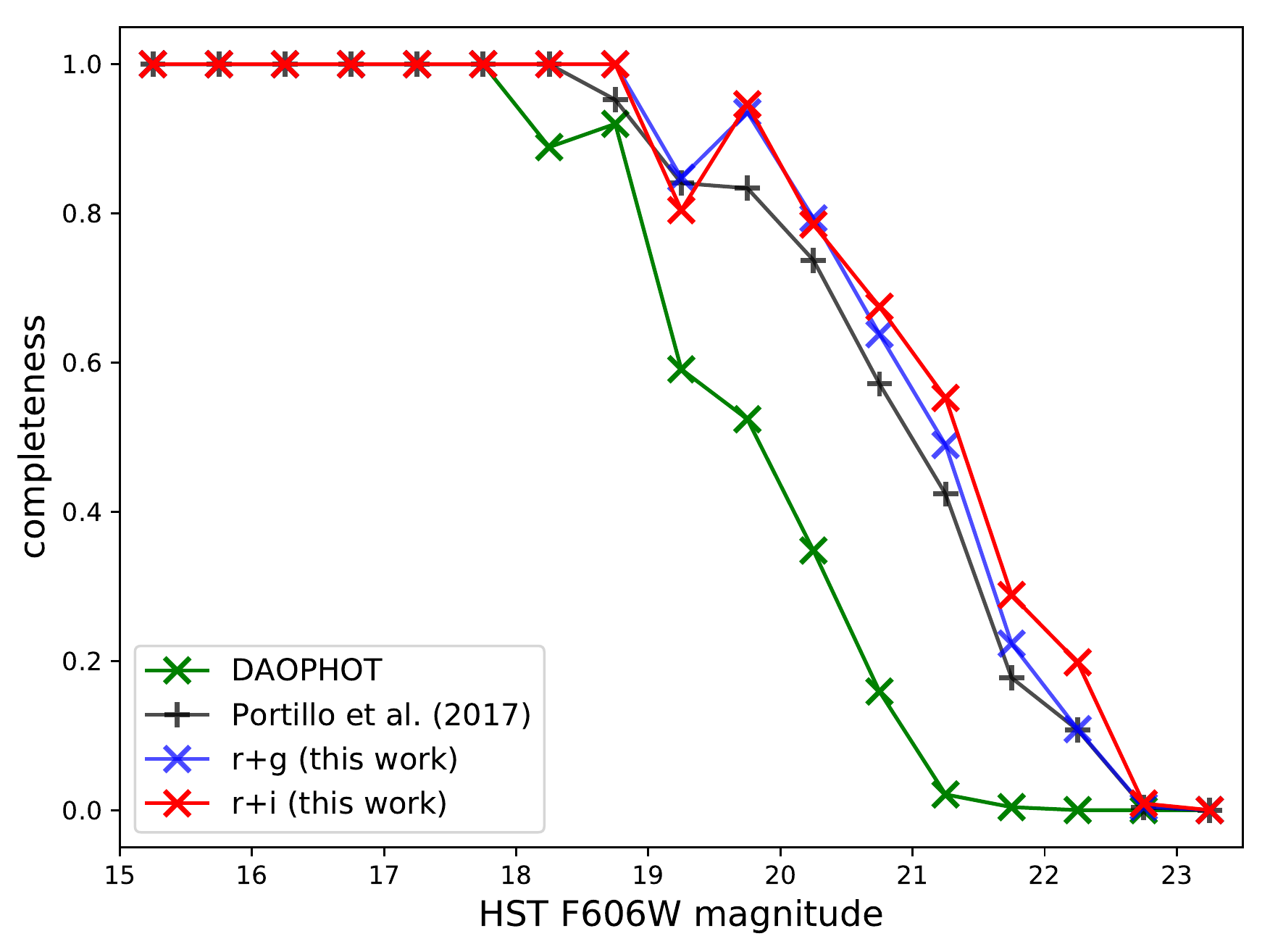}
    \includegraphics[width=0.48\linewidth]{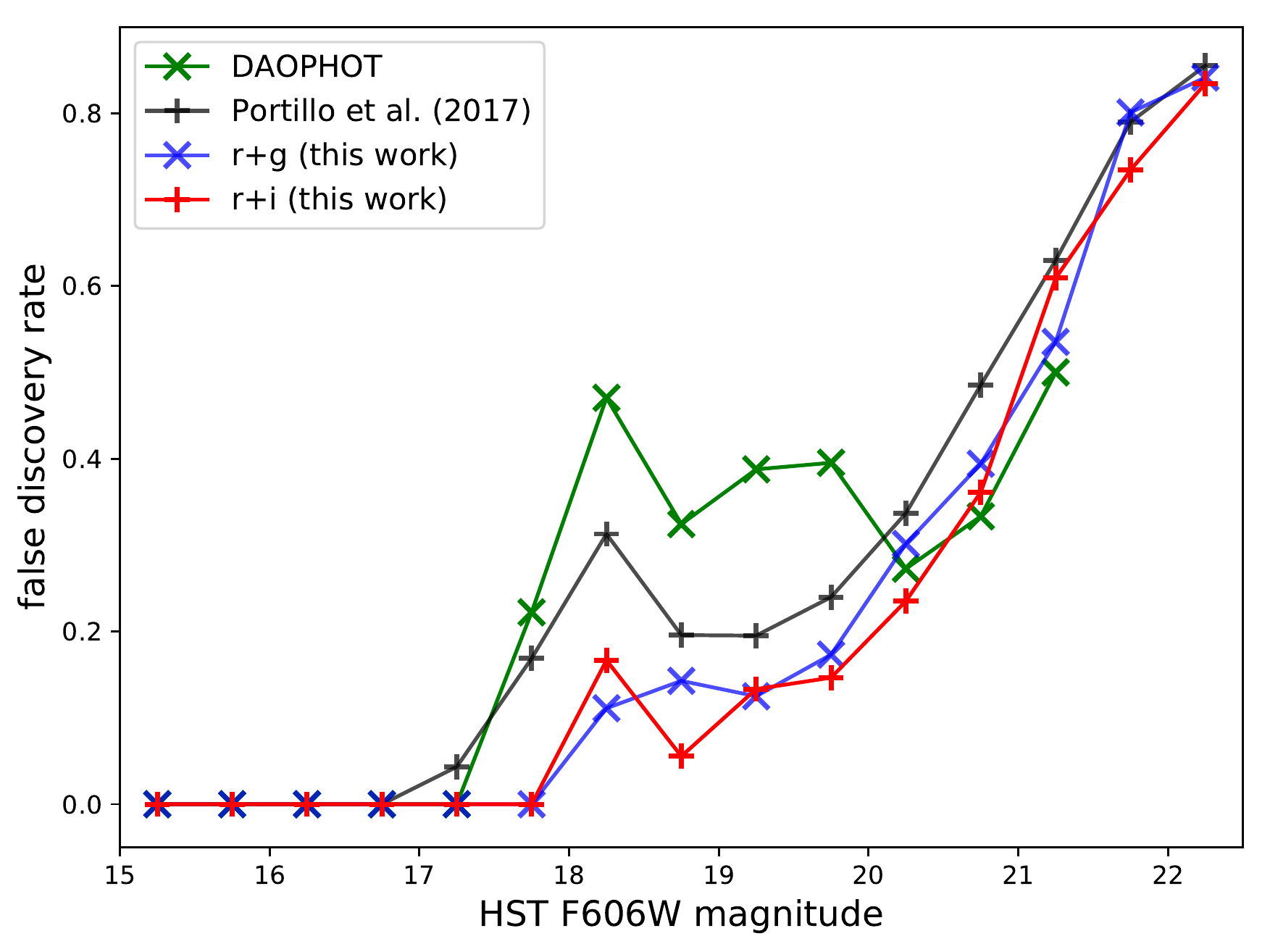}
    \caption{Completeness (left) and false discovery rate (right) as a function of magnitude for the two-band condensed catalogs (blue, red), the single-band catalog ensemble (black), and the DAOPHOT catalog (green). Completeness and false discovery rate are determined by comparing to the HST catalog of the same region, which we treat as ``ground truth" down to F606W$\sim 23$.}
    \label{fig:m2_comp_fdr}
\end{figure*}

Figure \ref{fig:m2_comp_fdr} compares the completeness and false discovery rates of the DAOPHOT catalog (black) and single-band catalog ensemble from \cite{PORTILLO_17} (green). The single-band probabilistic catalog goes more than a magnitude deeper than DAOPHOT while maintaining a lower false discovery rate. Also plotted are the completeness/false discovery rates from the joint $r+g$ (blue) and $r+i$ (red) condensed catalogs. Multiband catalog fits further improve the completeness by several tenths of a magnitude while yielding lower false discovery rates. These results are roughly consistent with both the SNR estimates and multiband mock tests from \S 3. The improved deblending of the two-band condensed catalogs can be seen clearly between 18\ts{th} and 20\ts{th} magnitude. In this range, over $40\%$ of all DAOPHOT catalog sources are false positives. The single-band probabilistic catalog from \cite{PORTILLO_17} achieves a false discovery rate between $20\%$ and $30\%$ in this range, while our two-band condensed catalogs are consistently under $20\%$.

\begin{figure*}
    \centering
    \includegraphics[width=0.48\linewidth]{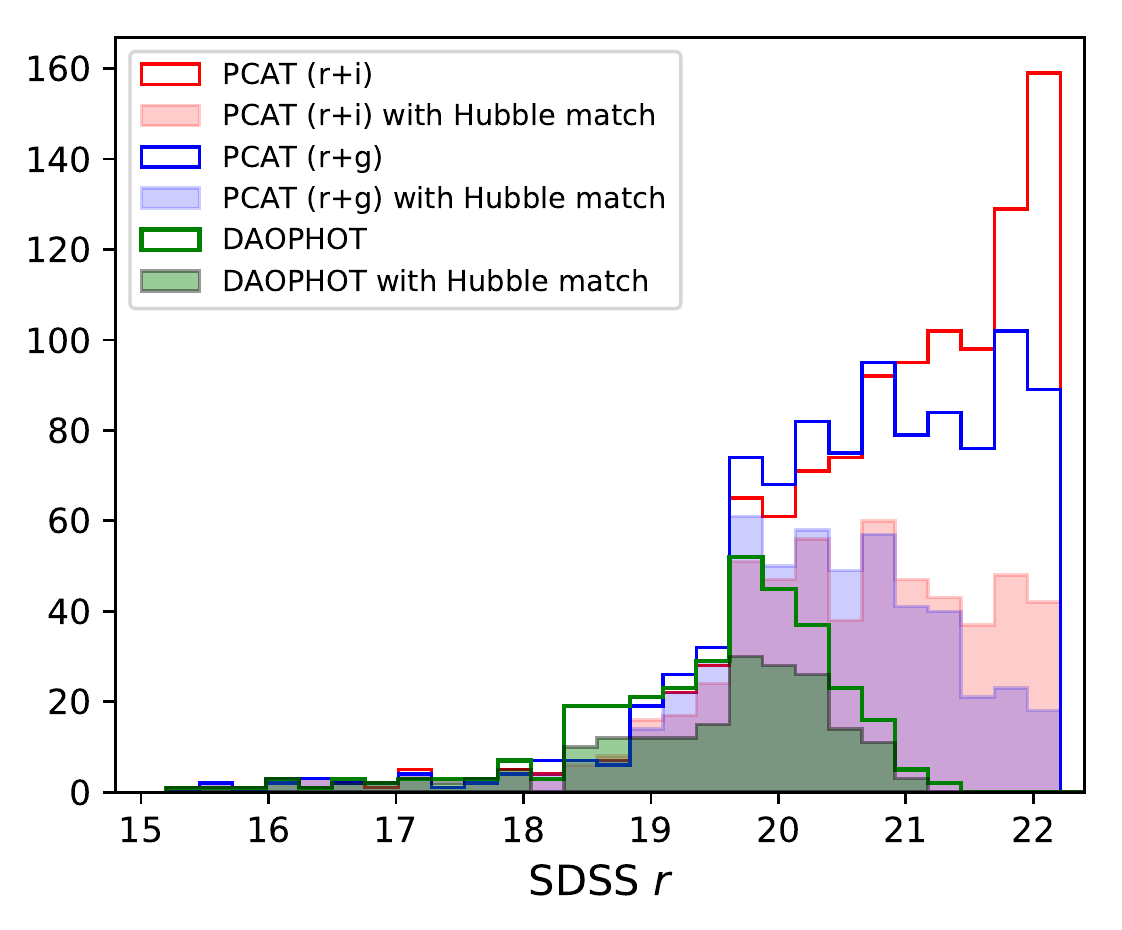}
    \includegraphics[width=0.48\linewidth]{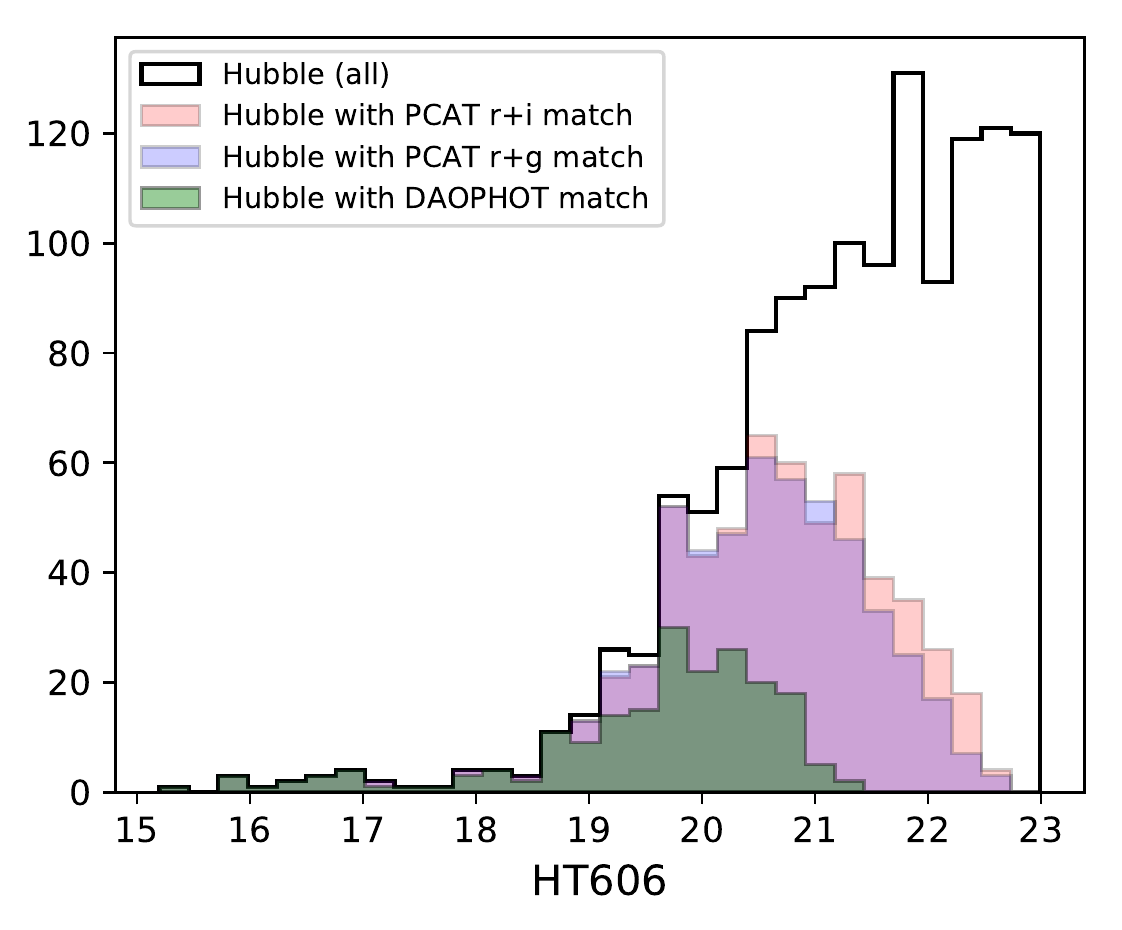}
    \caption{Source magnitude histograms for Hubble, PCAT, and DAOPHOT catalogs of M2. \textbf{Left:} PCAT and DAOPHOT catalogs binned by SDSS $r$-band. Shaded histograms show the distribution of PCAT/DAOPHOT sources with a corresponding match in Hubble catalog. \textbf{Right:} Hubble catalog binned by F606W magnitude. Shaded histograms show distribution of Hubble sources with corresponding match in PCAT condensed catalogs and DAOPHOT catalog.}
    \label{fig:m2_maghist}
\end{figure*}

Figure \ref{fig:m2_maghist} shows the binned magnitude histograms for Hubble, PCAT, and DAOPHOT catalog sources. It is clear that on the faint end PCAT detects many more sources that match with Hubble catalog sources than DAOPHOT does. The same improvements in completeness and false discovery seen in Figure \ref{fig:m2_comp_fdr} are also reflected by these magnitude distributions.

\begin{figure}
    \centering
    \includegraphics[width=\linewidth]{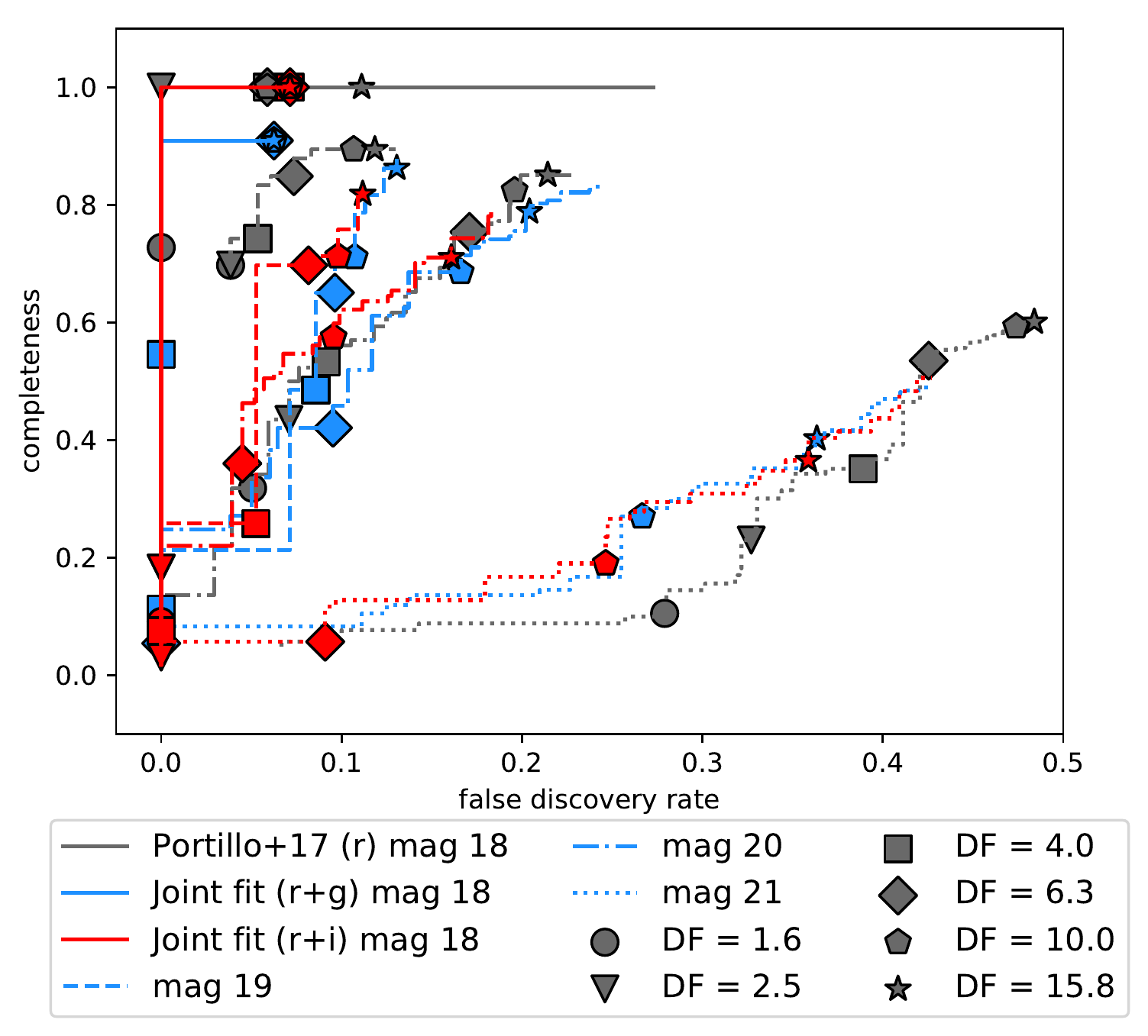}
    \caption{Completeness--false discovery tradeoff for the single-band condensed catalog from \cite{PORTILLO_17} (gray lines) and for the two-band $r+i$ (red) and $r+g$ (blue) condensed catalogs obtained in this work, evaluated for different $r$-band magnitudes. In place of a discriminative threshold as is commonly employed in ROC curves of binary classifiers, we impose successive cuts on the degradation factor to trace a curve in completeness--false discovery space. Several degradation factor cut values on each curve are denoted by symbols. Better performance lies closer to the upper left corner of the plot.}
    \label{fig:roc_curve}
\end{figure}

A useful quantity for comparing catalogs is the degradation factor, which is defined in \cite{PORTILLO_17} as the ratio of reported uncertainty in flux for a given source and expected uncertainty for an isolated source with the same flux. Sources with large degradation factors are those with large uncertainties, which are more likely to be heavily contaminated by neighbors or to be artifacts. As such, making successive cuts on the degradation factor is one way to describe the completeness--false discovery trade-off for our cataloger. Figure \ref{fig:roc_curve} shows completeness--false discovery rate curves for four magnitude bins between $r=17.5$ and 21.5, with each curve reporting values for one- and two-band condensed catalogs. For the magnitude bin centered on $r=21$, both $r+i$ and $r+g$ outperform the single-band condensed catalog -- one can tell that these are better because their ROC curves are closer to the upper left portion of the completeness--false discovery plane, which is ideal. For $r=20$, $r+i$ is better than single-band, but the difference is smaller. Both multiband fits are worse at $r=19$, and at $r=18$ the $r+g$ fit is worse. This decrease in performance may arise from the oversplitting of bright sources. Our estimates of completeness and false discovery come from one $100\times 100$ pixel SDSS field of view, so the calculated ROC curves are also affected by sampling noise.

\section{Discussion}

Multiband probabilistic cataloging provides a robust and flexible framework for point-source inference. By performing a simultaneous fit on data in multiple bands, one benefits from higher S/N (as reflected through the likelihood evaluation) along with the opportunity to use color information for more constraining power. Through a number of mock data tests, we demonstrate that these factors enhance the quality of the resulting catalog, in terms of both population characteristics and in disentangling the emission from highly covariant sources. 
Performing probabilistic cataloging on observational data requires careful treatment of systematics in the data. Working with crowded field data, we determine that errors from subpixel astrometric miscalibrations have a large impact on the quality of the catalog fit given the current implementation of PCAT. As shown with mock data in \S 3.2, cross-band astrometric miscalibration at the level of $10^{-2}$ pixels can lead to oversplitting of the brightest sources. While some fine-tuning helped keep oversplitting at a minimum in the case of two-band data from M2, applications to larger-scale data sets will require a more general approach to coordinate calibration. This might involve fitting for cross-band miscalibrations during burn-in using postage stamp regions around the brightest sources, or as part of an MCMC proposal that perturbs all of the sources in the fit by some small hyperparameter offset. We defer a proper treatment of this issue to future work.

In the case where probabilistic cataloging is performed using observations from different instruments, the systematics from each instrument should be handled carefully. More sophisticated corrections that account for multiple camera orientations, exposures, and sensitivities will be needed on a case-by-case basis to get a high-quality joint catalog fit. Another consideration will be the size and quality of the PSFs from different instruments. One might encounter matching degeneracies in which multiple sources identified in one band are within the PSF of an individual source from another band. For example, a report on the Massive Young Star Forming Complex in Infrared and X-ray (MYStIX) project \citep{MYSTIX_SRC} notes that cross-matching young stars is complicated by different PSF FWHMs and positional uncertainties in the case of off-axis \emph{Chandra} sources. By forward-modeling into all bands and fitting simultaneously, PCAT should go a long way in addressing these issues, but application-specific issues will remain.    

Absorption features and other phenomena may be relevant in certain applications of probabilistic cataloging, obscuring sources in certain passbands but not in others. A higher overall S/N means that sources that are fainter in certain bands may be easier to recover than otherwise. However, the crucial element to the success of PCAT in these situations is the formalism of the color prior. The color priors used in this work are broad and relatively unimposing, and they primarily help by constraining the parameter space explored by the Markov chains. One could, however, establish a number of spectral templates that place stronger constraints on well-understood objects. These templates could be drawn, for example, from stellar libraries. In this type of application, one would want to sample both within individual templates and across different templates. The Bayesian element of PCAT allows one to exploit previous knowledge of the spectra for various astrophysical objects in such a way that they can be used to perform inference on properties such as stellar type. Unlike the Naive Bayes classifier, a common algorithm used for classification tasks, PCAT does not require conditional feature independence, an assumption ill-suited for many applications in astronomy \citep{MYSTIX}. Tuning the color prior may help to identify sources whose models are well constrained, but it should be emphasized that restricting the color prior makes it more difficult for the model to identify sources that are spectrally unique. 

While this work focuses on multiband photometry, the framework of PCAT permits a wide range of modeling choices. Rather than fit spectral information, one could fit multiple observations from time series data, using prior information about the time variability of certain types of sources to detect and model such populations. As another example, one could employ a cluster hierarchical model in regions where the source number density is spatially dependent. This flexibility stems from the fact that PCAT uses a  hierarchical model.

\subsection{Scaling PCAT for Large-scale Astronomical Surveys}

This implementation of multiband probabilistic cataloging uses various computational and algorithmic optimizations to improve convergence. With half an hour of CPU time, it recovers similar results as the implementation from \cite{PORTILLO_17} does in 12 CPU-days, a speedup factor of over 500. This brings PCAT significantly closer to being a feasible pipeline for near-future surveys. 
Graphics processing units (GPUs) have been considered as a way to speed up probabilistic cataloging. However, many of the operations that would benefit from speed improvements are nontrivial to parallelize. For example, when sources are added or removed from an image, they need to be modified sequentially. There are modern CPUs that do this type of sequential operation efficiently. Another consideration is speed of matrix multiplication, which we use heavily in model evaluation (for subpixel shifts). Our matrix problem has a small internal dimension of 10, causing standard libraries (e.g. CUDA) to run at low GPU utilization. A modern Intel CPU (e.g. Sky Lake) with dual AVX-512 fused multiply-add (FMA) engines can perform 64 floating-point operations per cycle per core. Our PCAT code nearly saturates that theoretical limit even with a small internal dimension. Because we have a much higher utilization on CPUs versus GPUs (10$\times$ or more), we currently obtain better performance on CPUs. Future versions of PCAT that allow for a spatially varying PSF might have a much larger internal dimension in the matrix multiplication and scale better on GPUs. 

A crucial feature of any cataloger used for large-scale surveys is the ability to model galaxies, which are present in most fields in the sky. Galaxies with parameterized shapes (e.g. Sersic profile) can be incorporated into PCAT's forward model; however, the higher dimensionality required to model galaxies makes sampling more challenging. Furthermore, such parameterizations may not be sufficient for galaxies with complicated morphologies. Future development of PCAT will focus on constructing efficient galaxy proposals.

The challenges of astronomical cataloging are tied to a number of more general problems within statistics. Through our prior that penalizes additional sources, we employ a form of model regularization, in which models are selected based on goodness of fit and model complexity. By sampling across models, we calculate relative model evidences (i.e., Bayes factors) by the relative abundance of samples from each model. Explicitly calculating the (normalized) model evidences is often computationally intractable in high-dimensional problems. Approximate methods exist such as variational inference (VI), which uses optimization techniques to maximize an evidence lower bound. VI excels at handling large data sets, where one wishes to quickly explore multiple models, but generally underestimates the variance of the posterior density (\cite{BLEI}, \cite{REGIER}). 

Another alternative is to implement Hamiltonian reversible jump Monte Carlo, as has been done in \cite{HMC}. In their framework, transdimensional moves have Hamiltonian Monte Carlo moves added at the beginning and/or end. The transdimensional move may then not require much tuning, as the Hamiltonian moves will guide the proposal to higher likelihoods. Preliminary tests of fixed-dimensional HMC on cataloging have been done with promising results.

\section{Conclusion}
In this work we have implemented a multiband probabilistic cataloger that performs simultaneous forward modeling across bands and uses color information to improve point-source inference. Looking first at mock SDSS data, we show that two- and three-band fits produce catalog ensembles with completeness 0.4 and 0.6 mag deeper than single-band PCAT respectively, along with correspondingly lower false discovery rates. In \S 4 we demonstrate that both increased S/N and color information help disentangle highly blended sources, outperforming single-band analyses down to source separations of 0.5 pixels. When applied to SDSS data from the globular cluster M2, the two-band joint fits detect sources $\sim 0.4$ mag deeper than single-band probabilistic catalogs, while maintaining lower false discovery rates. Both single and multiband probabilistic cataloging methods vastly outperform standard maximum likelihood cataloging methods. We identify that when mishandled, errors in astrometric calibration and the PSF can produce oversplitting of the brightest sources. 

Through Bayesian forward modeling, probabilistic cataloging provides a robust framework to enhance source detection and deblending using multiband photometric datasets. Given recent improvements in time performance and sampling efficiency, PCAT should be seriously considered for further development as a pipeline equipped to address the challenges posed by near-future surveys.  

\acknowledgments
R.M.F is supported by the California Institute of Technology and was also supported by the Harvard-Smithsonian Center for Astrophysics. S.K.N.P. acknowledges support from a Sir James Lougheed Award of Distinction as well as the DIRAC Institute in the Department of Astronomy at the University of Washington. The DIRAC Institute is supported through generous gifts from the Charles and Lisa Simonyi Fund for Arts and Sciences, and the Washington Research Foundation. T.D. acknowledges support from MIT's Kavli Institute as a Kavli postdoctoral fellow.

We thank Aneta Siemiginowska, Vinay Kashyap, and Josh Speagle for useful comments throughout the course of this work. 

The computations in this paper were run on the Odyssey
cluster supported by the FAS Division of Science, Research Computing Group at Harvard University. This research has made use of NASA's Astrophysics Data System.

Funding for the Sloan Digital Sky Survey IV has been provided by the Alfred P. Sloan Foundation, the U.S. Department of Energy Office of Science, and the Participating Institutions. SDSS-IV acknowledges
support and resources from the Center for High-Performance Computing at
the University of Utah. The SDSS website is \url{www.sdss.org}.

SDSS-IV is managed by the Astrophysical Research Consortium for the 
Participating Institutions of the SDSS Collaboration including the 
Brazilian Participation Group, the Carnegie Institution for Science, 
Carnegie Mellon University, the Chilean Participation Group, the French Participation Group, Harvard-Smithsonian Center for Astrophysics, 
Instituto de Astrof\'isica de Canarias, The Johns Hopkins University, 
Kavli Institute for the Physics and Mathematics of the Universe (IPMU) / 
University of Tokyo, Lawrence Berkeley National Laboratory, 
Leibniz Institut f\"ur Astrophysik Potsdam (AIP),  
Max-Planck-Institut f\"ur Astronomie (MPIA Heidelberg), 
Max-Planck-Institut f\"ur Astrophysik (MPA Garching), 
Max-Planck-Institut f\"ur Extraterrestrische Physik (MPE), 
National Astronomical Observatories of China, New Mexico State University, 
New York University, University of Notre Dame, 
Observat\'ario Nacional / MCTI, The Ohio State University, 
Pennsylvania State University, Shanghai Astronomical Observatory, 
United Kingdom Participation Group,
Universidad Nacional Aut\'onoma de M\'exico, University of Arizona, 
University of Colorado Boulder, University of Oxford, University of Portsmouth, 
University of Utah, University of Virginia, University of Washington, University of Wisconsin, 
Vanderbilt University, and Yale University.

\software{Lion \citep{PORTILLO_17}, crowdsource \citep{Schlafly_2018}, NumPy, Matplotlib, CBLAS, SciPy}

\appendix
\section{Multiband Catalog Priors}\label{A}
\subsection{Flux/Color Priors}
Consider a catalog with $N$ sources observed in $k$ bands. The catalog may be represented as $C = \lbrace(x,y,f_1,f_2,...,f_k)\rbrace_{n=1}^N$, as in \S 2.1. In the single-band case, one would simply include a prior $\pi(f_1)$ on the flux of each source in that band. Trivially, one could extend this to multiple bands, imposing the same type of power law prior to each flux and using those in conjunction with one another. Instead of this, we impose a single reference band flux prior ($r$-band in this work) along with color priors relating the remaining bands to the reference band. This takes the following form:
\begin{equation}
\label{multiband_flux_prior}
\pi(\vec{f}) = \pi(f_1) \times \prod_{i=2}^k \pi(s_{i}),
\end{equation}
in which the color $s_{i} = -2.5\log_{10}(f_i/f_1)$. The color prior allows us to make use of the fact that the fluxes of a source observed in multiple bands are not independent. We choose a Gaussian prior on color $\pi(s_{i}) \sim \mathcal{N}(\mu_{i}, \sigma_{i})$. The means $\mu_{i}$ and Gaussian widths $\sigma_{i}$ depend on how well individual colors are constrained within certain astronomical populations. In practice, relatively wide Gaussian widths are used, only disfavoring sources with extremely atypical colors (e.g., $|r-i|>4$). When new sources are created in birth moves, these color priors can be useful. Rather than draw independent flux samples for a new source, we choose to draw one flux $f_1$ and then choose other fluxes $f_i$ by sampling from the colors $\pi(s_{i})$ and using those to calculate $f_i$. This results in more efficient proposals where birthed sources are drawn from a reasonable stellar locus.  
\subsection{Birth/Death Parsimony Prior}
To avoid overfitting our catalog model, PCAT institutes a prior on $N$, the number of sources. This parsimony prior, also referred to in the literature as a regularization prior, penalizes each additional source in the model. For an idealized Gaussian problem in one band, adding a source corresponds to an average improvement in log-likelihood of the maximum likelihood solution by 3/2 (1/2 per degree of freedom). This is a result of Wilks's theorem, which states that for a likelihood ratio $\Lambda$ between hypotheses $\Theta$ and $\Theta_0$, the quantity $2\log\Lambda$ will asymptotically be chi-squared distributed as $\chi^2$ with degrees of freedom equal to the difference in dimensionality between $\Theta$ and $\Theta_0$ \citep{WILKS}. The parsimony prior that counteracts this overfitting effect is $\pi(N) \propto \exp(-\frac{3}{2} N)$. Because we add parameters to our model for each additional band, the prior is modified to 
\begin{equation}
\pi(N) \propto \exp\left({-\frac{N}{2}(2+n_{bands})}\right).
\label{parsimony_prior}
\end{equation}
While a Poisson prior on the mean number of sources $N$ may be more consistent with our Bayesian hierarchical model, it is also less punishing of new sources. Tests using the Poisson prior produced a catalog ensemble with many sources that did not affect bright sources and had low S/N. If one specifically wanted to estimate the faint-end source population, the parsimony prior would not be appropriate. However, the focus of this work is deblending, and the parsimony prior in Equation \eqref{parsimony_prior} helps cut down on CPU time requirements because the penalty for new sources is stronger.

With these changes in effect, our prior is proportional to the following:
\begin{equation}\label{prior}
\pi(\vec{\theta}) \propto \pi(N)\prod_{n=0}^N \pi(x_n,y_n)\pi(\vec{f_n}) = \pi(N)\prod_{n=0}^N \pi(x_n,y_n)\pi(f_{n,1})\prod_{i=2}^k \pi(s_{n,i}).
\end{equation}
\section{Proposals and Acceptance Fractions}\label{B}
\subsection{Flux/Position Proposals}
In the case of multiple bands, we have some degree of freedom regarding how we can propose source fluxes. When proposing new sources, our implementation draws the reference band flux from a power law, while the remaining fluxes are calculated through fair draws on Gaussian color priors. To perturb the fluxes of existing sources, we draw proposals directly on all fluxes, such that our proposal distribution resembles the eventual stationary distribution on individual fluxes.

The size of position-shifting proposals also depends on source fluxes and is modified in the multiband case. While the optimal step size for position proposals is calculated using quantities relevant to the image, we make a modification that scales for more degrees of freedom per source. Fixing $N_{src}$ somewhere on the order of $N_{max}$ (2000 in this work), 
\begin{equation}
\sigma_x \propto \frac{1}{N_{src}(2+n_{bands})} \frac{1}{f_1}.
\end{equation} 
To maintain detailed balance,
$f_1 = \max(f_{1,current}, f_{1,proposed})$, which is conserved between a proposed step and its inverse. 

\subsection{Merges/Splits}
In merge/split proposals, we would like to conserve the flux of our original source/sources. To some extent, we would like to conserve the color of our original source as well, in the sense that we would like to propose sources with reasonable colors. If the fraction of original source flux is drawn randomly as $F_k \sim \text{Unif}[0,1]$ for each band separately, the two sources produced from a split are likely to have unreasonable colors, in which case our prior on color will suppress the proposal acceptance factor. Our proposal generates sources with colors perturbed from the original source color. In the reference band, we draw $\rho \sim \text{Unif}[0,1]$ and calculate $F_1 = (f_{min}/f)+\rho(1-2f_{min}/f)$. This is designed so that as $f$ approaches $2f_{min}$ (the lowest permissible to be a candidate for a split), $F_1$ goes to 0.5, and for $f \gg f_{min}$, $F_1$ approaches $\rho$. For each band $k$ other than the reference band, we generate a delta color $\Delta s \sim \mathcal{N}(0, \sigma_s)$, from which the flux fraction is calculated as
\begin{equation}
	F_k = \frac{\exp\left(\frac{\Delta s}{\kappa}\right)F_1}{1-F_1+\exp\left(\frac{\Delta s}{\kappa}\right)F_1}
\end{equation}
where $\kappa = \frac{2.5}{\ln 10} \approx 1.08.$ One source receives fraction $F_k$ of the original source flux in band $k$, while the other receives $1-F_k$. The colors of the two sources are then
\begin{align}
	s_{1,k} &= \kappa\ln\frac{F_k}{F_1} \\
    s_{2,k} &= \kappa\ln \frac{1-F_k}{1-F_1}.
\end{align}

The acceptance factor for merges and splits also needs to be modified in the case of multiple bands. Consider the case of a proposal to split one source into two smaller sources. The acceptance ratio for a split can be expressed as
\begin{align}
\label{merge_split_alpha}
\alpha_{split} &= \frac{\pi_1(x,y,f_1,\lbrace{s_i\rbrace}_{i=2}^n)\pi_2(x,y,f_1,\lbrace{s_i\rbrace}_{i=2}^n)}{\pi_0(x,y,f_1,\lbrace{s_i\rbrace}_{i=2}^n)} \frac{\mathcal{J}}{q(\Delta x, \Delta y, \lbrace{F_i\rbrace}_{i=1}^n)} \\
 &= \frac{2\pi k^2}{A} \frac{\pi_1(f_1)\pi_2(f_1)}{\pi_0(f_1) q(F_1)} \prod_i \frac{\pi_1(s_i)\pi_2(s_i)}{\pi_0(s_i) q(\Delta c_i)}\mathcal{J} \label{alpha_split},
\end{align}
Here $\mathcal{J}$ represents the Jacobian associated with our change of variables and $q$ is the transition kernel from which proposals are drawn. The Jacobian in $n$ bands is calculated to be
\begin{equation}
\mathcal{J} = \prod_{i=1}^n \frac{\kappa}{F_i(1-F_i)}
\end{equation} The factor at the front of Equation \eqref{alpha_split} accounts for the kick range $k$ at which the two split sources are separated, along with $A$, the area of the region being evaluated.

The corresponding acceptance factor associated with mergers is simply the reciprocal of the split likelihood factor, which maintains detailed balance across proposals. While the split/merge proposal used here is not fully optimized, it does maintain detailed balance and is sufficient for our purposes.

\subsection{Background Proposals}
Floating background levels in a catalog fit can be difficult because the background is highly covariant with sources in a given image. In the case where the background is allowed to float as a free parameter in the catalog model, the optimal background proposal step size is given by $\sigma_{b} \sim \sigma_{pixel}\sqrt{1/N_{pix}}$ in units of photoelectrons. In this expression, $\sigma_{pixel}$ is the pixel variance that depends on the data and gain, and $N_{pix}$ is the total number of pixels in the region being perturbed. Most observational datasets give estimates of background for their images, though in the case of crowded fields these estimates are not always accurate because of the covariance mentioned above.    

In cases when one wants to understand physical properties of an astrophysical background, floating the background level is an important element of the analysis. In the optical case, however, we are able to obtain a reasonable fit so long as the background is constrained at the $\sim 10\%$ level. There is still covariance between background and astronomical sources in optical data, but exploring this covariance is not the focus of this analysis, and so for our purposes we choose to fix the background in our main runs. Future analysis of optical data with PCAT will likely consider this in more detail, especially because the background cannot be assumed to be spatially uniform over larger regions of sky.
\section{Sparse Field Validation}
To validate our work, we run multiband PCAT on a $500\times500$ pixel region from run 8151, camcol 4, field 63. The source number density in field 8151 is $\sim 10^3$ smaller than that of the crowded field examined in this work. As such, the catalog from SDSS imaging pipeline \texttt{Photo} was sufficient as reference and detects 19 stars and 31 galaxies in the region. We again look at bands $r$, $i$, and $g$. Since PCAT models emission with point sources, we only focus on how well PCAT models the stars in our image. Because we look at a larger region than before, we use a median filter to capture the $\sim 10$ ADU background variation across each image. While similar oversplitting issues persist, the multiband runs detect all 19 sources. 

\section{Simultaneous Fit of Observations vs. Co-addition}
The properties of catalog sources are commonly estimated through bootstrapping techniques in which single-band catalogs are treated as resampled versions of the full, multiband catalog. However, this approach does not benefit from the combined S/N of multiple observations -- catalog sources must be detectable in \emph{each} band. For this reason, many cataloging pipelines employ the co-addition of observations in which images are stacked on top of each other and the resulting image is fed into the cataloger. In this section we show that, given an identical PSF across all bands, a simultaneous fit always has a greater delta log-likelihood than a fit of a stacked image, with equality between log-likelihoods when source fluxes are the same across all observations.

The log-likelihood for a single image is
\begin{equation}
\log \mathcal{L} = \sum_{i\in pixels} -\frac{(d_i-fp_i)^2}{2\sigma^2},
\end{equation}
where $f$ is the model flux, $\sigma$ is the pixel noise, and $d_i$ and $p_i$ are the data and PSF at pixel $i$, respectively. The best-fit flux $\hat{f}$ can be found by taking the derivative with respect to flux:
\begin{equation}
0 = \frac{\partial \log \mathcal{L}}{\partial f} = \sum_{i\in pixels} \frac{(d_i - fp_i)p_i}{\sigma^2} \rightarrow \hat{f}=\frac{\Sigma_i d_ip_i}{\Sigma_i p_i^2}
\end{equation}
The delta log-likelihood between the true flux and zero flux (i.e., no source) is
\begin{equation}
\Delta \log \mathcal{L} = \sum_{i\in pixels} \frac{d_i^2 - (d_i-f^{*}p_i)^2}{2\sigma^2}
\end{equation}
If $f^*$ is indeed the true flux, then $d_i = f^*p_i + \delta_i$ with $\delta_i \sim \mathcal{N}(0, \sigma^2)$, so in expectation,
\begin{equation}
\langle\Delta \log \mathcal{L}\rangle = \Big\langle\sum_{i\in pixels} \frac{d_i^2 - (d_i-f^{*}p_i)^2}{2\sigma^2}\Big\rangle = \Big\langle\sum_{i\in pixels} \frac{f^{*^2}p_i^2 + 2f^*p_i\delta_i}{2\sigma^2}\Big\rangle = \frac{f^{*^2}}{2\sigma^2}\sum_{i\in pixels}p_i^2.
\end{equation}

Consider detecting a source using a simultaneous fit over images of different bands. The log-likelihood for each band adds
\begin{equation}
\langle\Delta \log \mathcal{L}\rangle_{multi} = \sum_{j\in images} \frac{f_j^{\star^2}}{2\sigma_{f,j}^2}.
\end{equation}

Now consider detecting a source using some stacking procedure. The method we will use in this example is inverse variance weighting, such that the stacked image has pixel noise level
\begin{equation}
S^2 = \Big(\sum_j\frac{1}{\sigma_j}\Big)^{-1},
\end{equation}
where we assume that the summation is over images to simplify notation. The stacked image itself is
\begin{equation}
D_i = \Big(\sum_j \frac{d_{ij}}{\sigma_j^2}\Big)S^2.
\end{equation}
In a stacked image, the fluxes get effectively averaged:
\begin{equation}
F^* = \Big(\sum_j\frac{f_j^*}{\sigma_{f,j}^2}\Big)\Big(\sum_j\frac{1}{\sigma_{f,j}^2}\Big)^{-1}
\end{equation}
with the average having an error
\begin{equation}
S_F^2 = \Big(\sum_j\frac{1}{\sigma_{f,j}^2}\Big)^{-1}.
\end{equation}
Therefore, the delta log-likelihood in the stacked image is 
\begin{align}
\langle\Delta \log \mathcal{L}\rangle_{stack} &= \frac{F^{*^2}}{2S_F^2} = \frac{1}{2}\Big(\sum_j\frac{f_j^*}{\sigma_{f,j}^2}\Big)^2\Big(\sum_j\frac{1}{\sigma_{f,j}^2}\Big)^{-1} = \frac{1}{2}\Big(\sum_j\frac{f_j^*}{\sigma_{f,j}}\Big)^2 S_F^2 \\
&= \frac{S_F^2}{2}\sum_j\sum_k\frac{f_j^* f_k^*}{\sigma_{f,j}^2\sigma_{k,j}^2}.
\end{align}
Now, we wish to show that
\begin{equation}
\sum_j\frac{f_j^{*^2}}{\sigma_{f,j}^2} \geq \frac{F^{*^2}}{S_F^2}.
\end{equation}
To do this, we first divide both sides by $S_F^2$ and calculate the difference between terms:
\begin{align}
\frac{1}{S_F^2}\sum_j\frac{f_j^{*^2}}{\sigma_{f,j}^2} - \frac{F^{*^2}}{S_F^4} &= \sum_j \frac{f_j^{*^2}}{\sigma_{f,j}^2}\Big(\sum_j \frac{1}{\sigma_{f,j}^2}\Big) - \sum_j \sum_k \frac{f_j^* f_k^*}{\sigma_{f,j}^2\sigma_{k,j}^2} \\
&= \sum_j \sum_k \frac{f_j^{*^2} - f_j^*f_k^*}{\sigma_{f,j}^2\sigma_{k,j}^2}
\end{align}
Because $j$ and $k$ iterate over the same images, we can say that by index interchange,
\begin{equation}
\sum_j\sum_k \frac{f_j^{*^2}}{\sigma_{f,j}^2\sigma_{k,j}^2} = \sum_j\sum_k \frac{f_k^{*^2}}{\sigma_{f,j}^2\sigma_{k,j}^2}
\end{equation}
so that we can write
\begin{align}
\frac{1}{S_F^2}\sum_j \frac{f_j^{*^2}}{\sigma_{f,j}^2} - \frac{F^{*^2}}{S_F^4} &= \sum_j\sum_k \frac{f_j^{*^2}-f_j^*f_k^*}{\sigma_{f,j}^2\sigma_{k,j}^2} \\
&= \sum_j\sum_k \frac{\frac{1}{2}f_j^{*^2}-f_j^*f_k^*+\frac{1}{2}f_k^{*^2}}{\sigma_{f,j}^2\sigma_{k,j}^2} \\
&= \sum_j\sum_k \frac{\frac{1}{2}(f_j^* - f_k^*)^2}{\sigma_{f,j}^2\sigma_{k,j}^2} \geq 0.
\end{align}
Equality only holds when $f_j^* = f_k^*$ $\forall j, k$, or when the flux is the same in all bands. Using this result, 
\begin{equation}
\langle\Delta \log \mathcal{L}\rangle_{stack} = \frac{F^{*^2}}{2S_F^2} \leq \sum_j\frac{f_j^{*^2}}{2\sigma_{f,j}^2} = \langle\Delta \log \mathcal{L}\rangle_{multi}.
\end{equation}
Thus, a simultaneous fit has a delta log-likelihood that is greater than or equal to that of the corresponding stacked image, with equality when the flux is the same in all bands. While more sophisticated stacking procedures exist, any stacking procedure will be lossy by construction.



\end{document}